\documentclass[twocolumn,showpacs,preprintnumbers,fleqn,amsmath,amssymb]{revtex4}

\usepackage{enumerate}
\usepackage{graphicx}
\usepackage{dcolumn}
\usepackage{bm}
\newtheorem {proposition}{Proposition}[section]
\newtheorem{lemma}{Lemma}[section]

\begin{document}

\preprint{IFT UwB /01/2002}

\title{Some Integrable Systems \\ in Nonlinear Quantum
Optics }

\author{Maciej Horowski, Anatol Odzijewicz and Agnieszka Tereszkiewicz}%
 \email{ horowski@alpha.uwb.edu.pl, aodzijew@labfiz.uwb.edu.pl,
tereszk@alpha.uwb.edu.pl}
\affiliation{Institute of Theoretical Physics\\
University in Bia{\l}ystok
\\Lipowa 41, 15-424 Bia{\l}ystok, Poland}

\date{\today}

\begin{abstract}
In the paper we investigate the theory of quantum optical systems. As an application we integrate
and describe the quantum optical systems which are generically related to the classical orthogonal
polynomials. The family of coherent states related to these systems is constructed and described.
Some applications are also presented.
\end{abstract}

\maketitle
\section{Introduction}

Quantum optics affords a big amount of very interesting physical phenomena having important
application at the same time. Our aim is to formulate the theory of these phenomena and elucidate
their connection with the theory of orthogonal polynomials. This allows us to use the last one for
the rigorous integration of some nonlinear quantum optical models describing the interaction of the
finite number of modes of electromagnetic field with nonlinear medium. Let us mention, that in
quantum optical literature, see e.g. \cite{A-I}, \cite{B-C}, \cite{J}, \cite{Kar 1}, \cite{Kar 2}
the solutions of the models of this type are usually approached by approximative or semiclassical
methods.

In Section 2 we deduce from natural and  not restrictive assumptions, the general form, see
(\ref{Aal}), of the Hamiltonian $\mathbf H_I$  describing the interaction of the finite number of
modes of electric field with the mater. Later on, in Section 3, we investigate the quantum
reduction method which allows us to describe the quantum optical systems by the use of the theory
of orthogonal polynomials (see \cite{A}, \cite{A-G}, \cite{O-H-T}). We also show that the reduced
system are related to some quantum algebras, see relations (\ref{Abi}). These algebras  were
investigated in \cite{Odz 1}, where their relations to the theory of special and $q$-special
functions was shown.

Having  spectral measure of the interaction Hamiltonian, which is for example the case if the model
under consideration is related to the classical orthogonal polynomials  or their $q$-deformation,
see \cite{O-H-T}, \cite{H}, we can introduce spectral coherent states. They are direct
generalizations of Glauber coherent states (corresponding to the Hermite polynomials) and squeezed
states. In Section 4, we show that spectral coherent states admit the holomorphic representation
for the Hamiltonians under consideration which supplements the spectral and Fock representations.
This simplifies remarkably the calculation of many important physical characteristics of the
described system.

The spectral coherent states should have some physical meaning which needs the deeper
understanding. In any case, they give the link of quantum  optical systems with complex analytic
and symplectic geometry.  This opens the application of coherent states method, investigated in
\cite{Odz 1}, \cite{Odz 2} to the problems of the theory of quantum optics.

In Section 5. we give complete solution of the quantum systems, see (\ref{ca}), related to the
classical orthogonal polynomials.

Finally, in Section 6, we present the physical interpretation of the Hamiltonian given by
(\ref{Aan}) as the parametric modulator, which includes as special cases such quantum optical
systems as nondegenerate parametric amplifier and the frequency up-converter, see \cite{W-M}.

At the end we express our conviction that the proposed method will be helpful in better
understanding of quantum optical problems.

\section{Quantum electromagnetic field in nonlinear medium}
\renewcommand{\theequation}{2.\arabic{equation}}
\setcounter{equation}{0}

 Nonlinear optics deals with phenomena that occur as a consequence of the
modification of the optical properties of a material system in the presence of light. Practically,
only laser light is sufficiently intensive to produce the measurable effects. By an optical
nonlinearity we mean that the dipole moment per unit volume, or polarization $\vec{P}$, of a
material system depends in nonlinear way, upon the strength of the applied electromagnetic field.
As many authors \cite{B-C}, \cite{P-L} we assume that $\vec{P}$ depends only on the electric part
$\vec{E}$ of the electromagnetic field $(\vec{E},\vec{B})$ i.e. $\vec{P}\equiv\vec{P}[\vec{E}]$. We
assume moreover that this dependence is a functional one. Thus, in most general case we can write
\begin{eqnarray}
 \lefteqn{ \vec{P}[\vec{E}](t,\vec{r})=\displaystyle{\varepsilon_0\sum\limits_{N=0}^\infty
 \int dt_1\; d^3\vec{r}_1 \ldots dt_N\;d^3\vec{r}_N}\label{Aa}}\\
 &&\!\!\!\!\times\vec{T}_{(N)}\left(t,\vec{r},t_1,\vec{r}_1,...,t_N,\vec{r}_N\right)
 \left(\vec{E}\left(t_1,\vec{r}_1\right),...,\vec{E}\left(t_N,\vec{r}_N\right)
\right),\nonumber
 \end{eqnarray}
where the vector valued $N$-linear map $\vec{T}_{(N)}$, is called in optical literature $N$-th
response tensor of the medium \cite{B-C}.

The time-invariance principle, which says that the dynamical properties of the system are assumed
to be unchanged by a translation of the time origin, leads to
\begin{eqnarray}
\lefteqn{\vec{T}_{(N)}(t,\vec{r},t_1,\vec{r}_1,\ldots,t_N,\vec{r}_N)}\nonumber\\
&&=:\vec{R}_{(N)} (\vec{r},t_1-t,\vec{r}_1,t_2-t,\vec{r}_2,\ldots,t_N-t,\vec{r}_N).\label{Ab}
\end{eqnarray}
The interaction of the electromagnetic field $(\vec{E},\vec{B})$ with a nonlinear medium
characterized by polarization $\vec{P}$ can be described by the source-free Maxwell equations
 \begin{eqnarray}\label{Ac}
  &&\!\!\!\!\nabla\times\vec{E}=-\;\frac{\partial}{\partial  t}\vec{B},\vspace{0.5ex}\nonumber\\
  &&\!\!\!\!\nabla\times\vec{B}=\mu_0\;\frac{\partial}{\partial  t}(\varepsilon_0\vec{E}+
  \vec{P}[\vec{E}]),\vspace{1ex}\\
  &&\!\!\!\!\nabla \cdot  (\varepsilon_0\vec{E}+\vec{P}[\vec{E}])=0,\nonumber\vspace{4ex}\\
  &&\!\!\!\!\nabla \cdot \vec{B}=0\;.\nonumber\vspace{8ex}
 \end{eqnarray}
Therefore the divergence of the Poynting vector $\frac{1}{\mu_0}\vec{E}\times\vec{B}$ takes the
form
 \begin{eqnarray}
 \frac{1}{\mu_0}\nabla\cdot
 (\vec{E}\times\vec{B})&=&-\frac{\partial}{\partial
 t}\left(\frac{1}{2\mu_0}\vec{B}^2+\frac{\varepsilon_0}{2}\vec{E}^2\right)\nonumber\\
 &&-\vec{E}\cdot
 \frac{\partial\vec{P}}{\partial t}[\vec{E}],\label{Ad}
 \end{eqnarray}
where the quantity
 \begin{eqnarray}\label{Ae}
 u_0:=\frac{1}{2\mu_0}\vec{B}^2+\frac{\varepsilon_0}{2}\vec{E}^2
 \end{eqnarray}
is the energy density of the free  electromagnetic field. Analogously, we define the interaction
energy density $u_1$ by the equation
 \begin{eqnarray}\label{Af}
 \frac{\partial u_I}{\partial t}:=\vec{E}\cdot\frac{\partial
 \vec{P}}{\partial t}[\vec{E}]\,.
 \end{eqnarray}
The energy density
 \begin{eqnarray}\label{Ag}
 u(t,\vec{r}):=u_0(t,\vec{r})+u_1(t,\vec{r})
 \end{eqnarray}
determines the Hamiltonian $H$ of our system
 \begin{eqnarray}\label{Ah}
 H=H_0+H_1,
 \end{eqnarray}
where
 \begin{eqnarray} \displaystyle{H_0=\int u_0
(t,\vec{r})\,d^3\vec{r}}\label{Ai}
 \end{eqnarray}
is the Hamiltonian of the free electromagnetic field and the Hamiltonian
 \begin{eqnarray}\label{Aj}
\displaystyle{H_1=\int u_1(t,\vec{r}) \,d^3\vec{r}}\,
 \end{eqnarray}
describes the interaction of electric field $\vec{E}$ with the medium under consideration.

In order to obtain an explicit formula for $u_1$ let us consider the electromagnetic field
potential $\vec{A}$
 \begin{eqnarray}
 \vec{A}(t,\vec{r})=\left.\sum\limits_\lambda \int d^3\vec{k}\right[&&\!\!\!\!\!\!\!\vec{e}_{k,\lambda}A_\lambda(\vec{k})
e^{i(\omega_k t-\vec{k}\cdot\vec{r})}\nonumber\\
&&\!\!\!\!\!\!\!\!\!\left.+\vec{e}\;^*_{k,\lambda}A^*_\lambda(\vec{k}) e^{-i(\omega_k
t-\vec{k}\cdot\vec{r})} \right] ,\label{Ak}
 \end{eqnarray}
 expressed in terms of Fourier modes, where the index  $\lambda\in
\{1,2\}$ labels the polarization of the field, which is described by the pair of unit vectors
$\vec{e}_{k,1}$ and $\vec{e}_{k,2}$ orthogonal to the wave vector $\vec{k}$ (we choose the Coulomb
gauge $\nabla\cdot\vec{A}=0$). Here we do not specify the form of the dispersion relation, so we
assume that $\omega_k$ is any function of $|\vec{k}|$.
 In this gauge  we have
 \begin{equation}\label{Am}
 \vec{E}=-\frac{\partial}{\partial t}\vec{A}\quad \quad \quad\quad \quad \quad\vec{B}=
\nabla\times \vec{A}.
 \end{equation}
Let us introduce the following simplifying notation
 \begin{eqnarray}\label{An}
&&\!\!\!\!\!\!\!\vec{e}\;^\sigma_{k,\lambda}:=\left\{
  \begin{array}{l}
  -i\omega_k\:\vec{e}\;_{k,\lambda}\quad \quad \quad\;\,\textrm{for} \;\;\sigma=1\\
  i\omega_k\:\vec{e}\;^*_{k,\lambda}\quad \quad \quad\quad \,
\textrm{for} \;\;\sigma=-1
  \end{array}\right.\\
 &&\!\!\!\!\!\!\!\label{Ao}
  A^\sigma_{\lambda}(\vec{k}):=\left\{
  \begin{array}{l}
  A_{\lambda}(\vec{k})\quad \quad \quad\quad \;\; \textrm{for} \;\;\sigma=1\\
  A^*_{\lambda}(\vec{k})\quad \quad \quad\quad \;\; \textrm{for} \;\;\sigma=-1
  \end{array}\right.\quad.
  \end{eqnarray}
We have now
 \begin{equation} \label{Ap}
 \vec{E}\left(t,\vec{r}\right)=\sum_{\lambda,\sigma} \int
\vec{e}\;^\sigma_{k,\lambda} \;A^\sigma_{\lambda}(\vec{k}) \;e^{\sigma
i(\omega_kt-{\vec{k}}\cdot{\vec{r}})}\,d^3\vec{k}
 \end{equation}
and, therefore, (\ref{Aa}) becomes
\begin{widetext}\vspace{-4ex}\begin{eqnarray}
\displaystyle{\vec{P}[\vec{E}]\left(t,\vec{r}\right)=\sum\limits_{N=0}^\infty
 \;\sum\limits_{\sigma_1,\lambda_1}\ldots\sum\limits_{\sigma_N,\lambda_N}\;\;
 \int}&&\!\!\!\!\!\! d^3\vec{k}_1\ldots d^3\vec{k}_N\vspace{1ex}\nonumber\;\;\vec{\chi}_{(N)}\left(\vec{r},\sigma_1,\vec{k}_1,\omega_{k_1},\ldots,\sigma_N,
 \vec{k}_N,\omega_{k_N}\right)\left( \vec{e}\;^{\sigma_1}_{k_1,\lambda_1},\ldots, \vec{e}\;^
 {\sigma_N}_{k_N,\lambda_N}\right)\\
 &&\times\; e^{it\sum\limits_{r=1}^N \sigma_r\omega_{k_r}}A^{\sigma_1}_{\lambda_1}(
 \vec{k_1})\ldots A^{\sigma_N}_{\lambda_N}(\vec{k_N }),\label{Ar}
 \end{eqnarray}
where $\vec{\chi}_{(N)}\left(\vec{r},\sigma_1,\vec{k}_1,\omega_{k_1},\ldots,\sigma_N,
  \vec{k}_N,\omega_{k_N}\right)$, is  the $N$-th susceptibility
tensor \cite{B-C} defined  by
 \begin{eqnarray}\label{As}\lefteqn{
 \vec{\chi}_{(N)} \!\!\left( \!\vec{r},\sigma_1,\vec{k}_1,\omega_{k_1},...,\sigma_N,
 \vec{k}_N,\omega_{k_N}\!\right) \!\!= \!\!\int \!\! \vec{R}_{(N)}\left(\vec{r},\tau_1,\vec{r}_1,...,\tau_N,
 \vec{r}_N\right)\vspace{0.5ex}
 \,e^{i\sum\limits_{s=1}^N\sigma_s\omega_{ks}\tau_s}\,
 e^{-i\sum\limits_{s=1}^N\sigma_s\vec{k}_s\cdot\vec{r}_s}
 \,d\tau_1\,d^3\vec{r}_1 ... d\tau_n\,d^3\vec{r}_N.}
 \end{eqnarray}
Inserting  (\ref{Ar}) into (\ref{Af}) we  find up to additive constant that
 \begin{eqnarray}
 \lefteqn{\displaystyle{u_1(t,\vec{r})=\sum\limits_{N=0}^\infty\;\sum\limits_{\sigma_0,\lambda_0}
 \;\sum\limits_{\sigma_1,\lambda_1}...\sum\limits_{\sigma_N,\lambda_N}\int\! d^3\vec{k}_1...d^3\vec{k}_N\vspace{1ex}}\nonumber\;\;\vec{e}\;^{\sigma_0}_{k_0,\lambda_0}\cdot\vec{\chi}_{(N)}\left(\vec{r},
 \sigma_1,\vec{k}_1,\omega_{k_1},...,\sigma_N,
 \vec{k}_N,\omega_{k_N}\right)\left( \vec{e}\;^{\sigma_1}_{k_1,\lambda_1},\ldots, \vec{e}\;^
 {\sigma_N}_{k_N,\lambda_N}\right)}\\
 &&\quad \quad \quad \quad \quad \quad \quad \quad \quad \quad \quad \quad
 \quad \quad \quad \times\; \frac{\sum\limits_{r=1}^N \sigma_r\omega_{k_r}}
 {\sum\limits_{s=0}^N \sigma_s\omega_{k_s}}\;e^{it\sum\limits_{s=0}^N\sigma_s\omega_{k_s}}
 e\;^{-i\sigma_0\vec{k}_0\cdot\vec{r}}\;A^{\sigma_0}_{\lambda_0}(\vec{k_0})
 A^{\sigma_1}_{\lambda_1}(\vec{k_1})\ldots A^{\sigma_N}_{\lambda_N}(\vec{k_N }).\label{At}
 \end{eqnarray}\vspace{-3ex}\end{widetext}

In the quantization procedure the classical quantities $ A^{\sigma}_{\lambda}(\vec{k})$ in
(\ref{Ak}) are replaced by the operators
 \begin{equation}\label{Au}
 {\mathbf a}^\sigma_{k,\lambda}:=\left\{
  \begin{array}{l}
    {\mathbf a}^{}_{k,\lambda}\quad \quad \quad\;\;\textrm{for} \;\;\sigma=1\\
    {\mathbf a}^*_{k,\lambda}\quad \quad \quad \;\; \textrm{for}
\;\;\sigma=-1
  \end{array}\right.,
 \end{equation}
which   satisfy the  commutation relations of a free quantum field:
 \begin{eqnarray}\label{Av}
 \left[ {\mathbf a}^\sigma_{\lambda,k}\;,{\mathbf
 a}^{\sigma'}_{\lambda',k'}\right]=\delta_{\lambda\lambda'}
 \delta(k-k') \frac{\sigma}{2}(1-\sigma\sigma').
 \end{eqnarray}
The products $ A^{\sigma_0}_{\lambda_0}(\vec{k}_0)\ldots A^{\sigma_N}_{\lambda_N}(\vec{k}_N)$ in
(\ref{Ar}) and (\ref{At}) are moreover, replaced by the normally ordered products of corresponding
operators, i.e. by  $:{\mathbf a}^{\sigma_0}_{k_0,\lambda_0}\ldots {\mathbf
a}^{\sigma_N}_{k_N,\lambda_N}:\;$.

In order to obtain the Hamiltonian it is enough to insert (\ref{Au}) into (\ref{At}) and then
(\ref{At}) into (\ref{Aj}). With the Hamiltonian of the free electromagnetic field
\begin{equation}\label{Aw}
 {\mathbf H}_0=\sum_{\lambda}\int d^3\vec{k}\;\omega_k\;{\mathbf
a}^*_{k,\lambda}{\mathbf a}_{k,\lambda}^{}
 \end{equation}
 we obtain
 \begin{widetext}\vspace{-5ex}
 \begin{eqnarray}
 \displaystyle{{\mathbf H}={\mathbf H}_0+\sum\limits_{N=0}^\infty
 \sum\limits_{\sigma_0,\lambda_0}...\sum \limits_{\sigma_N,
 \lambda_N}\int}&&\!\!\!\!\!\!\!\! d^3\vec{k}_0... d^3\vec{k}_N \;:\mathbf a^{\sigma_0}_{k_0,\lambda_0}
 ... \mathbf
 a^{\sigma_N}_{k_N,\lambda_N}: e^{it\sum\limits_{s=0}^N
 \sigma_s\omega_{k_s}}\vspace{0.5ex}\nonumber\\
 &&\!\!\!\!\!\!\!\!\!\times\;\vec{e}\;^{\sigma_0}_{k_0,\lambda_0}\cdot\vec{\Theta}_{(N)}\left(\sigma_0,\vec{k}_0,
\omega_{k_0}, \sigma_1,\vec{k}_1,\omega_{k_1},...,\sigma_N,\vec{k}_N,\omega_{k_N}\right)
 \left(\vec{e}\;^{\sigma_1}_{k_1,\lambda_1},...,
 \vec{e}\;^{\sigma_N}_{k_N,\lambda_N}\right)\vspace{0.5ex}\label{Ax}\label{Az}\label{Ay},
 \end{eqnarray}
where
\begin{eqnarray}\label{Aaa}
 \displaystyle{\vec{\Theta}_{(N)}\!\left(\!\sigma_0,\vec{k}_0,\omega_{k_0},\sigma_1,\vec{k}_1,
 \omega_{k_1}...,\sigma_N,\vec{k}_N,\omega_{k_N}\!\right)\!:=
 \frac{\sum\limits_{r=1}^N \sigma_r\omega_{k_r}}
 {\sum\limits_{s=0}^N \sigma_s\omega_{k_s}}}
 \displaystyle{\int  \vec{\chi}_{(N)}\!\left(\!\vec{r},\sigma_1,\vec{k}_1,\omega_{k_1},...,\sigma_N,
 \vec{k}_N,\omega_{k_N}\!\right)\!e^{-i\sigma_0\vec{k}_0\cdot\vec{r}}\;d^3\vec{r}.}
 \end{eqnarray}
Using the commutation relations (\ref{Av}) one can prove that
 \begin{equation}\label{Aab}
 e^{-i{\mathbf H}_0t}\;{\mathbf
a}^{\sigma}_{k,\lambda}\;e^{i{\mathbf H}_0t}=e^{i\sigma \omega_k t}\;{\mathbf
a}^{\sigma}_{k,\lambda}\;.
 \end{equation}
Hence the Hamiltonian (\ref{Ah}) becomes
 \begin{equation}\label{Aac}
 {\mathbf H}= {\mathbf H}_0+ e^{-i{\mathbf H}_0t}\;{\mathbf H}_I
\;e^{i{\mathbf H}_0t},
 \end{equation}
where due to (\ref{Ax})
 \begin{eqnarray}
 \lefteqn{\displaystyle{{\mathbf H}_I=\sum\limits_{N=0}^\infty
\sum\limits_{\sigma_0,\lambda_0}\ldots\sum\limits_{\sigma_N,
 \lambda_N}\int d^3\vec{k}_0\ldots d^3\vec{k}_N:{\mathbf a}^{\sigma_0}_{k_0,\lambda_0}\ldots
 {\mathbf
 a}^{\sigma_N}_{k_N,\lambda_N}:}}\nonumber\\
&&\;\;\;\;\;\;\;\;\;\;\;\;\;\;\;\;\;\;\;\;\;\;\;\;\;\;\;\;\;\;\;\;\;\;\;\;\;\; \times
\vec{e}\;^{\sigma_0}_{k_0,\lambda_0}\cdot\vec{\Theta}_{(N)}\left(\sigma_0,\vec{k}_0,
 \omega_{k_0},\sigma_1,\vec{k}_1,\omega_{k_1}, \ldots,\sigma_N,\vec{k}_N,\omega_{k_N}\right)\left(
 \vec{e}\;^{\sigma_1}_{k_1,\lambda_1},\ldots,
 \vec{e}\;^{\sigma_N}_{k_N,\lambda_N}\right)\label{Aad}
 \end{eqnarray}\end{widetext}
does not depend on time, and therefore the solution of the Schr\"{o}dinger equation\vspace{-1ex}
\begin{equation}\label{Aae}
i\;\frac{\partial}{\partial t}|\psi(t)\rangle={\mathbf H}|\psi(t)\rangle
\end{equation}
is given by\vspace{-1ex}
\begin{equation}\label{Aaf}
|\psi(t)\rangle=e^{-i{\mathbf H}_0t}e^{-i{\mathbf H}_It}\;|\psi(0)\rangle.
\end{equation}
\vspace{-1ex}The operator
\begin{equation}\label{Aag}
\mathbf U_0(t):=e^{-i\mathbf H_0t}
\end{equation}
is the   free electromagnetic field evolution operator. The operator
\begin{equation}\label{Aag}
\mathbf U_I(t):=e^{-i\mathbf H_It}
\end{equation}
is the evolution operator of the system in the interaction picture.

For the real models in quantum optics one assumes that the system
 under consideration contains a finite number of modes of electric
field (see \cite{A-I}, \cite{J}, \cite{Kar 1}, \cite{P-L} ). This means that the label $(\vec{k},
\lambda)$ in (\ref{Aad}) and (\ref{Aw}) takes a finite number of values
\begin{equation}\label{Aah}
(\vec{k},\lambda)\equiv j\in\{0,1,\ldots,M\}
\end{equation}
and the integrals are reduced to finite sums over $j$:
\begin{eqnarray}
 {\mathbf H}_0&\!\!=\!\!&\sum_j \omega_j {\mathbf a}^*_j {\mathbf a}_j\label{Aai}\\
{\mathbf H}_I&\!\!=\!\!&\sum\limits_{N=0}^\infty
  \sum\limits_{\sigma_0,j_0}...\sum\limits_{\sigma_N,j_N}:{\mathbf a}^{\sigma_0}_{j_0}... {\mathbf
 a}^{\sigma_N}_{j_N}:\label{Aaj}\\
  && \times\,\vec{e}\;^{\sigma_0}_{j_0}
 \cdot\vec{\Theta}_{(N)}\left(\sigma_0,\omega_{j_0},
 ...,\sigma_N,\omega_{j_N}\right) \left(
 \vec{e}\;^{\sigma_1}_{j_1},...,\vec{e}\;^{\sigma_N}_{j_N}\right). \nonumber
 \end{eqnarray}
In this case  the Hamiltonian (\ref{Aaj}) can be transformed into the form which is  more useful
for our aims. It is defined by the exchange of the normal ordering of the annihilation and creation
operators into the one which we will call boson-number ordering in the sequel.

In order to define the \textbf{boson-number ordering} let us introduce the following notation for
creation and annihilation operators
\begin{equation}\label{Aak}
  {\mathbf a}^l:=\left\{
   \begin{array}{l}
   {\mathbf a}^l\quad\quad\quad\quad \textrm{for} \;\;l=1,2,\ldots\\
   1\quad\quad \quad\quad \; \textrm{for} \;\;l=0\\
   ({\mathbf a}^*)^{-l}\quad\quad \:\textrm{for}
\;\;l=-1,-2,\ldots
   \end{array}\right..
\end{equation}
A product of $m$ annihilation  and $n$ creation operators being in the same mode, is said to be
boson-number ordered if it is of the form $P({\mathbf a}^*{\mathbf a}^{}){\mathbf a}^{m-n}$, where
$P$ is a polynomial.

Changing the normal ordering in each term of the Hamiltonian (\ref{Aaj}) to the boson-number
ordering   $P({\mathbf a}_0^*{\mathbf a}_0^{},\ldots,{\mathbf a}^*_M {\mathbf a}_M^{})\;{\mathbf
a}_0^{l_0}\ldots {\mathbf a}_M^{l_M}\;$ where $P$ is a polynomial of $M+1$ variables and collecting
the terms with the same factor ${\mathbf a}_0^{l_0}\ldots {\mathbf a}_M^{l_M}$  we obtain
 \begin{equation}\label{Aal}
\mathbf H_I=\sum_{l_0,\ldots,l_M\in{\Bbb Z}}\;g_{l_0,\ldots,l_M}\left({\mathbf a}^*_0 {\mathbf
a}_0^{},\ldots,{\mathbf a}^*_M
 {\mathbf a}_M^{}\right){\mathbf a}_0^{l_0}\ldots {\mathbf
a}_M^{l_M},
 \end{equation}
where $g_{l_0,\ldots ,l_M}$ are   functions of $(M+1)-$variables dependent on $\vec{\Theta}_{(N)}$.
The
 Hamiltonian ${\mathbf H}_I$ is a symmetric operator if
\begin{eqnarray}
\lefteqn{[g_{l_0,\ldots,l_M}\left({\mathbf a}^*_0{\mathbf a}_0^{},\ldots,{\mathbf a}^*_M {\mathbf
a}_M^{}\right)]^*\nonumber }\\
&&=g^*_{-l_0,\ldots,-l_M}\left({\mathbf a}^*_0{\mathbf a}_0^{}-l_0,\ldots,{\mathbf a}^*_M {\mathbf
a}_M^{}-l_M\right).\label{Aam}
\end{eqnarray}
In the next sections we restrict our considerations to the Hamiltonians of the form
\begin{eqnarray}
{\mathbf H}_I&=&h\left({\mathbf a}^*_0{\mathbf a}_0^{},\ldots,{\mathbf a}^*_M
{\mathbf a}_M^{}\right)\nonumber\\
&&+g\left({\mathbf a}^*_0{\mathbf a}_0^{},\ldots,{\mathbf a}^*_M{\mathbf a}_M^{}\right){\mathbf
a}_0^{l_0}\ldots {\mathbf a}_M^{l_M} +h.c.\label{Aan}
\end{eqnarray}
Such form of the Hamiltonian is strictly related to the theory of orthogonal polynomials. The
physical interpretation of this Hamiltonian is given in the Paragraph \textbf{6.A}.
\section{Reduction of the Hamiltonian}
\renewcommand{\theequation}{3.\arabic{equation}}
\setcounter{equation}{0}

In this section, we briefly describe the decomposition  of the Hilbert space $\cal H$ spanned by
elements of the orthonormal Fock basis
\begin{widetext}
\begin{equation}\label{Aax}
{\cal B}_F=\left\{\left|n_0,\ldots,n_M\right\rangle:=\frac{1}{\sqrt{n_0!\ldots n_M!}}\left({\mathbf
a}_0^*\right)^{n_0}\ldots\left({\mathbf
a}^*_M\right)^{n_M}\left|0\right\rangle,\;\;n_0,\ldots,n_M\in{\Bbb N}\cup\{0\}\right\}
\end{equation}\end{widetext}
into invariant subspaces of the operators $\mathbf H_0$ and $\mathbf H_I$. The method of this
decomposition is presented in details in \cite{O-H-T}.  In such a way we obtain the reduction of
the Hamiltonian $\mathbf H$.

The  invariant subspaces of $\mathbf H_I$ are obtained in two steps. The first step is related to
some family of integrals of motion; the second one is related to a family of pseudo-vacuum vectors.

Let us start with a few definitions:
\begin{equation}\label{Aao}
{\mathbf A}:=g\left({\mathbf a}^*_0{\mathbf a}_0^{},\ldots,{\mathbf a}^*_M{\mathbf
a}_M^{}\right){\mathbf a}_0^{l_0}\ldots {\mathbf a}_M^{l_M}
\end{equation}
and
\begin{equation}\label{Aap}
{\mathbf A}_i^{}={\mathbf A}^*_i:=\sum_{j=0}^M\alpha_{ij}\;{\mathbf a}^*_j{\mathbf
a}_j^{},\;\;\;\;\;\;\;i=0,1,\ldots,M,
\end{equation}
where $\alpha=(\alpha_{ij})$ is a real $(M+1)\times (M+1)$-matrix satisfying the conditions
\begin{eqnarray}
\label{Aar}
 \det \alpha &\neq& 0, \\
 {\sum_{j=0}^M}\alpha _{ij}\;l_j &=& \delta _{0i}.\label{Aas}
\end{eqnarray}

The invertibility of the matrix $\alpha$ allows one to express the boson-number operators ${\mathbf
a}^*_i{\mathbf a}_i^{}$ by ${\mathbf A}_j$, which gives
 \begin{eqnarray}
\mathbf H_0=\sum_{j=0}^M \gamma_j \mathbf A_j
 \end{eqnarray}
with real constants $\gamma_j$ determined by the matrix $\alpha$. In particular we have
\begin{equation}\label{Aatt}
\gamma_0=\sum^M_{i=0}\omega_jl_j.
\end{equation}
Additionally
\begin{equation}\label{Aat}
{\mathbf H}_I={\mathbf H}_d({\mathbf A}_0,{\mathbf A}_1,\ldots,{\mathbf A}_M)+{\mathbf
A^{}}+{\mathbf A}^*
\end{equation}
with ${\mathbf H}_d$ uniquely determined by the function $h$ and the matrix $\alpha$. Using the
canonical commutation relations for creation and annihilation operators one obtain
\begin{eqnarray}
&&\!\!\!\!\!\!{\mathbf A}^{}{\mathbf A}^{*} ={\cal G}\left( {\mathbf A}_0,{\mathbf A}_1,\ldots
,{\mathbf
A}_M\right), \label{Aav}\\
&&\!\!\!\!\!\!{\mathbf A}^{*}{\mathbf A}^{} ={\cal G}\left( {\mathbf A}_0-1,{\mathbf A}_1,\ldots
,{\mathbf A}
_M\right)\label{Aau},\\
&&\!\!\!\!\!\![\mathbf A_0,\mathbf A]=-\mathbf A,\;\;\;\;\; [\mathbf A_0,\mathbf A^*]=\mathbf A^*,\\
&&\!\!\!\!\!\![\mathbf A_j,\mathbf
A]=0,\;\;\;\;\;\;\;\;\;\;\;j=1,\ldots,M,\\
&&\!\!\!\!\!\![\mathbf A_i,\mathbf A_j]=0,\;\;\;\;\;\;\;i,j=0,\ldots,M
\end{eqnarray}
with the nonnegative function ${\cal G}$ uniquely determined by $g$ and $\alpha$.

Direct calculations gives
\begin{equation}\label{Aaw}
[\mathbf A_j,\mathbf H_0]=[{\mathbf A}_j,{\mathbf H}_I]=0\;\;\;\;\;j=1,2,\ldots,M.
\end{equation}
which means that operators ${\mathbf A}_1,{\mathbf A}_2,\ldots,{\mathbf A}_M$ are integrals of
motion.

In order to reduce ${\mathbf H}_0$ and ${\mathbf H}_I$ to the common eigenspace of integrals of
motion let us notice that the operators ${\mathbf A}^*{\mathbf A},\;{\mathbf A^{}}{\mathbf
A}^*,{\mathbf A}_0,\ldots,{\mathbf A}_M$ are diagonal in the Fock basis ${\cal B}_F$. This, in
particular, means that each vector $\left|n_0,\ldots,n_M\right\rangle\in {\cal B}_F$ is the
eigenvector of the operators ${\mathbf A}_j,\;\;j=0,\ldots,M$, with eigenvalues given by:
 \begin{equation}\label{Aay}
 \lambda_j=\sum_{i=0}^M\alpha_{ji}\;n_i.
 \end{equation}
Moreover, the operators ${\mathbf A}_0,\ldots,{\mathbf A}_M$ form a system of commuting independent
observables.  In such a way we can use the sequences of eigenvalues
$(\lambda_0,\lambda_1,\ldots,\lambda_M)$ as a new parametrization
$\{\left|\lambda_0,\lambda_1,\ldots,\lambda_M\right\rangle\}$ of the Fock basis elements. So we
obtain
 \begin{equation}\label{Aaz}
 {\mathbf
 A}_j\left|\lambda_0,\lambda_1,...,\lambda_M\right\rangle=\lambda_j
 \left|\lambda_0,\lambda_1,...,\lambda_M\right\rangle,\;\;\;j=0,...,M.
 \end{equation}
Since $[{\mathbf A}_0,{\mathbf A}]=-{\mathbf A}$ then, from (\ref{Aau}) and (\ref{Aav}) we have
 \begin{eqnarray}
 \lefteqn{{\mathbf
 A}\left|\lambda_0,\lambda_1,\ldots,\lambda_M\right\rangle
 \nonumber}\\
 &&\!=\sqrt{{\cal G}\left( \lambda_0-1,\lambda_1,...
 ,\lambda_M\right)}
 \left|\lambda_0-1,\lambda_1,...,\lambda_M\right\rangle\!, \label{Aba}
 \end{eqnarray}
\begin{eqnarray}
 \lefteqn{ {\mathbf
 A}^*\left|\lambda_0,\lambda_1,\ldots,\lambda_M\right\rangle\nonumber}\\
 & &=\sqrt{{\cal G}\left( \lambda_0,\lambda_1,...
 ,\lambda_M\right)}\left|\lambda_0+1,\lambda_1,...,\lambda_M\right\rangle\label{Abb}.
 \end{eqnarray}

It is clear that the subspace ${\cal H}_{\lambda_1\ldots\lambda_M}$ of the Fock space $\cal H$
spanned by the eigenvectors $\left|\lambda_0,\lambda_1,\ldots,\lambda_M\right\rangle$ with fixed
$\lambda_1,\ldots,\lambda_M$ is ${\mathbf H}_0$ and ${\mathbf H}_I$-invariant and $dim{\cal
H}_{\lambda_1\ldots\lambda_M}=\infty$ if and only if  all $l_j$ in (\ref{Aao}) are nonnegative. The
problem of integration of the system (\ref{Aac}) is reduced to integration of the system described
by the reduced Hamiltonian
 \begin{eqnarray}\label{Abc}
 \mathbf H_{0,red}&:=&\gamma_0{\mathbf A}_0+\sum_{j=1}^M\gamma_j\lambda_j,\\
  \mathbf H_{I,red}&:=&{\mathbf H}_d\left({\mathbf
A}_0,\lambda_1,\ldots,\lambda_M\right)+{\mathbf A}+{\mathbf A}^*\label{Abcc}
 \end{eqnarray}
and therefore (up to additive constant)
 \begin{eqnarray}
 \mathbf H_{red}=\gamma_0{\mathbf A}_0+e^{-i\gamma_0\mathbf
A_0t}\;\,\mathbf H_{I,red}\;\,e^{i\gamma_0\mathbf A_0t}.
 \end{eqnarray}

Now we go to the next step of the reduction. In order to make it let us define the pseudo-vacuum
vector as  such vector $\left|\lambda_0,\lambda_1,\ldots,\lambda_M\right\rangle$ from the Fock
basis in ${\cal H}_{\lambda_1,\ldots,\lambda_M}$ which is annihilated by the operator ${\mathbf
A}$, i.e.
 \begin{eqnarray}\label{Abd}
{\mathbf A}\left|\lambda_0,\lambda_1,\ldots,\lambda_M\right\rangle=0
 \end{eqnarray}
or equivalently
 \begin{eqnarray}\label{Abe}
 {\cal G}(\lambda_0-1,\lambda_1,\ldots,\lambda_M)=0.
 \end{eqnarray}
In \cite{O-H-T} it was shown that the set $\{\lambda_{0,l}\}_{l=1}^K:=\{\lambda_0:\;{\mathbf
A}\left|\lambda_0,\lambda_1,\ldots,\lambda_M\right\rangle=0\}$ of the solutions of (\ref{Abd}) is
nonempty if in the definition (\ref{Aao}) any $l_i,\;\;i=0,1,\ldots,M$, is greater then zero.

Now, if for simplicity, we introduce the notation
 \begin{eqnarray}\label{Abf}
 |n\rangle&:=&|\lambda_{0,l}+n,\lambda_1,\ldots,\lambda_M\rangle,\nonumber\\
 b(n)&:=&\sqrt{{\cal G} (\lambda_{0,l}+n-1,\lambda_1, \ldots,
 \lambda_M)},\\
 {\mathbf N}&:=&{\mathbf A}_0-\lambda_{0,l}\;\;,\nonumber
 \end{eqnarray}
then
 \begin{eqnarray}
 {\mathbf N}|n\rangle&=&n|n\rangle,\nonumber\\
{\mathbf  A}|n\rangle&=&b(n)|n-1\rangle,\label{Abg}\\
{\mathbf  A}^*|n\rangle&=&b(n+1)|n+1\rangle\nonumber.
 \end{eqnarray}
Thus we obtain that the space
 \begin{eqnarray}\label{Abh}
 {\cal F}:=span\{\;|n\rangle,\;n=0,1,\ldots\}
 \end{eqnarray}
is the irreducible representation space for the algebra ${\cal A}_{red}$ generated by the operators
${\mathbf N},{\mathbf A}$ and ${\mathbf A}^*$, which satisfy the relations:
 \begin{eqnarray}
 [{\mathbf N},{\mathbf A}]&=&-{\mathbf A},\;\;\;\;[{\mathbf N},{\mathbf A}^*]
 ={\mathbf A}^*,\nonumber\\
 {\mathbf A}^*{\mathbf A}&=&b^2({\mathbf N}),\label{Abi}\\
{\mathbf A}{\mathbf A}^*&=&b^2({\mathbf N}+1).\nonumber
 \end{eqnarray}
 These algebras were investigated in \cite{Odz 2}. The question
when the dimension of $\cal F$ is finite or infinite was discussed in detail in \cite{O-H-T}. Here
we assume that $dim\,{\cal F}=\infty$. After restriction to $\cal F$, the Hamiltonians
 (\ref{Abc})  (\ref{Abcc}) belongs to ${\cal A}_{red}$ and take
the form (up to additive constant)
 \begin{eqnarray}
 \mathbf H_{0,red}&=&\gamma_0 \mathbf N,\\
{\mathbf H}_{I,red}&=&h({\mathbf N})+{\mathbf A}+{\mathbf A}^*\label{Abj},
 \end{eqnarray}
where $h({\mathbf N}):={\mathbf H}_d({\mathbf N}+\lambda_{0,l},\lambda_1,\ldots,\lambda_M)$. Thus
the operators $\mathbf H_{0,red}\,,\;\mathbf H_{I,red}\,$ and consequently $\mathbf H_{red}$ belong
to the algebra ${\cal A}_{red}$. In the Fock basis $\{\;|n\rangle,\;n=0,1,\ldots\}$ the operator
$\mathbf H_{I,red}$ assumes the three diagonal (Jacobi) form:
\begin{eqnarray}
\lefteqn{{\mathbf H}_{I,red}|n\rangle}\nonumber\\
&&=h(n)|n\rangle+b(n)|n-1\rangle+b(n+1)|n+1\rangle,\label{Abk}
 \end{eqnarray}
whereas  $\mathbf H_{0,red}$ is diagonal
\begin{eqnarray}
{\mathbf H}_{0,red}|n\rangle=\gamma_0\,n|n\rangle.
 \end{eqnarray}

 From now on we  restrict our consideration to the space $\cal F$.
In particular we restrict all  operators discussed above to $\cal F$ and  omit the index $red$ for
simplicity.

The evolution of the system given by (\ref{Aaf}) now takes the form
\begin{eqnarray}
|\psi(t)\rangle=e^{-i\gamma_0\mathbf N t}e^{-i\mathbf H_I t}|\psi(0)\rangle
\end{eqnarray}
and therefore for any operator $\mathbf F$ we have
\begin{eqnarray}
\lefteqn{\langle\psi(t)|\,\mathbf F\,\psi(t)\rangle\nonumber}\\
&&=\langle\psi(0)|\,e^{i\mathbf H_I t}e^{i\gamma_0 \mathbf N t }\mathbf F e^{-i\gamma_0 \mathbf N t
}e^{-i\mathbf H_I t}|\psi(0)\rangle.
\end{eqnarray}
For a special, but interesting case this formula simplifies. Namely,
\begin{eqnarray}
\lefteqn{\langle\psi(t)|\,f(\mathbf N )\,\psi(t)\rangle}\nonumber\\
&&=\langle\psi(0)|\,e^{i\mathbf H_I t} f(\mathbf N) e^{-i\mathbf H_I t}|\psi(0)\rangle\label{1},
\end{eqnarray}
\begin{eqnarray}
 \lefteqn{\langle\psi(t)|f( \mathbf A) \,\psi(t)\rangle}\nonumber\\
 &&=f(e^{-i\gamma_0t})\label{2}
\langle\psi(0)|\,e^{i\mathbf H_I t}f(  \mathbf A)\; e^{-i\mathbf H_I t}|\psi(0)\rangle,
\end{eqnarray}
\begin{eqnarray}
\langle\psi(t)|\,\mathbf H\,\psi(t)\rangle&=&\gamma_0 \langle\psi(0)|\,e^{i\mathbf H_I t}\mathbf N
e^{-i\mathbf H_I
t}|\psi(0)\rangle\nonumber\\
&&\label{3} +\langle\psi(0)|\,\mathbf H_I\psi(0)\rangle,
\end{eqnarray}
where $f$ is an analytic function.

The next section is devoted to the detailed study of the operator $\;\mathbf H_I\;$ and
1-parameter group $\;e^{-i\mathbf H_I t}\;$ generated by it.

\section{ Spectral and coherent states representations}
\renewcommand{\theequation}{4.\arabic{equation}}
\setcounter{equation}{0}

The operators $\mathbf{H}_I$ of the type (\ref{Abj}) are very well known in the theory of
orthogonal polynomials \cite{A}, \cite{A-G}, \cite{Ch}. They are symmetric in $\cal F$ and, by
(\ref{Abk}),  have a dense domain  which consists of finite linear combinations of elements of the
Fock basis. The deficiency indices of $\mathbf{H}_I$ are $(0,0)$ or $(1,1)$. One can prove that if
$\sum\limits_{n=1}^\infty\frac{1}{b(n)}=\infty$, then the operator $\mathbf{H}_I$ has deficiency
indices $(0,0)$, which is equivalent to its essential selfadjointness.

From now on we will assume that the deficiency indices of $\mathbf{H}_I$ are $(0,0)$. Hence
$\mathbf{H}_I$ admits a unique selfadjoint extension, which will be denoted by the same symbol.
Moreover, $\mathbf{H}_I$ has simple spectrum. This fact allows us to identify the  Fock space $\cal
F$ with the Hilbert space of square integrable functions $L^2({\Bbb R},\,d\sigma)$ of real variable
$\omega\in \Bbb R$. The measure $d\sigma$ is determined by the spectral measure $\mathbf dE$ of the
hamiltonian $\mathbf{H}_I$ and is defined by the formula
\begin{equation}
d\sigma (\omega):=\langle 0|{\mathbf dE}(\omega)0\rangle.\label{Ba}
\end{equation}
Additionally one can prove,  that polynomials $\{\omega^n\}_{n=0}^\infty$ form a linearly dense
subset in $L^2({\Bbb R},\,d\sigma)$. After the Gram-Schmidt orthonormalization of the basis
$\{\omega^n\}_{n=0}^\infty$ we obtain an orthonormal set $\{P_n\}_{n=0}^\infty$ in $L^2({\Bbb
R},\,d\sigma)$ called the orthonormal polynomial system. Notice that $deg\,P_n=n$.

The unitary isomorphism $\mathbf U:{\cal F}\longrightarrow L^2({\Bbb R},d\sigma)$ of Hilbert spaces
is given by
\begin{equation}\label{Bb}
\mathbf U |\psi\rangle:=\sum_{n=0}^\infty \langle n|\psi\rangle P_n.
\end{equation}
According to the spectral theorem and (\ref{Bb}) one has
\begin{equation}\label{Bd}
\left(\mathbf U\circ f(\mathbf H_I)\circ \mathbf U^{-1}\right)\psi(\omega)=f(\omega)\psi(\omega)
\end{equation}
for $\psi\in L^2({\Bbb R},d\sigma)$ and any measurable function $f$. By (\ref{Bb}) and (\ref{Bd})
the expression (\ref{Abk}) converts into the three-term recurrence formula
\begin{equation}\label{Be}
\omega P_n(\omega)=h(n)P_n( \omega)+b(n)P_{n-1}(\omega) +b(n+1)P_{n+1}(\omega)
\end{equation}
for the system of orthonormal polynomials $\{P_n\}_{n=0}^\infty$. So by the spectral theorem in the
notion of this orthonormal system, we have:
\begin{equation}\label{Bg}
\langle m|f(\mathbf H_I)\, n\rangle=\int f(\omega) P_m(\omega)P_n(\omega) \,d\sigma (\omega).
\end{equation}
In particular, for $f(\mathbf H_I)=\frac{1}{P_0^2}\mathbf H_I^k$ we obtain the moments $\mu_k$ of
the measure (\ref{Ba}):
\begin{equation}\label{Bh}
\mu_k:=\int \omega^k\,d\sigma (\omega)=\frac{1}{P_0^2}\langle 0|\;\mathbf H_I^k\,|0\rangle.
\end{equation}
Similarly, for $f(\mathbf H_I)=\frac{1}{P_0^2}|\mathbf H_I|^k$ we obtain the absolute moments
$|\mu_k|$ of (\ref{Ba}):
 \begin{equation}\label{Bi}
|\mu_k|:=\int |\omega|^k\,d\sigma (\omega)=\frac{1}{P_0^2}\langle 0|\;\;|\mathbf H_I|^k\;|0\rangle.
\end{equation}
For the case under consideration the moments $\{\mu_k\}_{k=0}^\infty$ determine $d\sigma$ in the
unique way \cite{A}.

From (\ref{Bd}) and (\ref{Bg}) one obtains that the evolution operator $e^{-i\mathbf H_It},\;
t\in\Bbb R$, in the Hilbert space $L^2({\Bbb R},d\sigma)$ is given by
\begin{equation}\label{Bj}
\left(\mathbf U\circ e^{-i\mathbf H_It}\circ \mathbf U^{-1}\right)\psi(\omega)=e^{-i\omega
t}\psi(\omega)
\end{equation}
and its mean value in the vacuum is realized by the characteristic function
\begin{equation}\label{Bk}
\widehat{\sigma}(t):=\int e^{-i\omega t}d\sigma(\omega)=\frac{1}{P_0^2}\left\langle 0\right|
e^{-i\mathbf H_I t}\left| 0\right\rangle
\end{equation}
of the measure $d\sigma$,  compare with \cite{F}.

After these preliminary remarks we will show that apart from realizations of the Hamiltonian
$\mathbf H_I$ in the Fock space $\cal F$ and in the Hilbert space $L^2\left({\Bbb R},
d\sigma\right)$ it is  useful and  natural to consider its realization in some Hilbert space  which
consists of square integrable holomorphic functions defined on an open subset of complex plane. To
do it let us first prove the following
 \begin{lemma}\label{BA}
Let us assume that absolute moments $|\mu|_n$ are finite for all $n\in {\Bbb N}\cup\{0\}$ and they
satisfy the condition
\begin{equation}\label{Bp}
\overline{\lim_{n\rightarrow\infty}}\frac{\sqrt[n]{|\mu|_n}}{n}=:\frac{1}{eR}<+\infty.
\end{equation}
Then there exists a maximal strip in $\Bbb C$, which is open, connected and invariant under the
one-parameter group of translations
\begin{equation}\label{Br}
T_tz:=z+t,\;\;\;t\in {\Bbb R},
\end{equation}
such that the characteristic function $\widehat{\sigma}(t)$ can be holomorphically extended to it.

The maximality of the strip means that $\widehat{\sigma}(t)$ cannot be extended to a larger set
with the same properties.
 \end{lemma}

{\it Proof} :  We prove firstly that characteristic function $\widehat{\sigma}$ is analytic on the
strip $|Im\, z|<R$. One has
\begin{equation}\label{Bt}
\frac{d^n}{dt^n}\,\widehat{\sigma}(t)=(-i)^n\int e^{-it\omega}\omega^nd\sigma(\omega)
\end{equation}
for $n=0,1,\ldots\;$. In order to prove (\ref{Bt}) one proceeds by induction. The equality
(\ref{Bt}) is valid for $n=0$. Let us assume that it is true for $n$, then
\begin{eqnarray}
{\frac{d^{(n+1)}}{dt^{(n+1)}}}\,\widehat{\sigma}(t)&=&\lim_{h\rightarrow\infty}
\int\frac{e^{-i(t+h)\omega}
-e^{-it\omega}}{h}\omega^n\;d\sigma(\omega)\nonumber\\
&=&\int\lim_{h\rightarrow\infty}\frac{e^{-ih\omega}-1}{h}e^{-it\omega}\omega^n
\;d\sigma(\omega)\nonumber\\
&=&-i\int e^{-it\omega}\omega^{n+1}d\sigma(\omega)\;\label{Bu}.
\end{eqnarray}
Since
 \begin{eqnarray}\label{Bw}
 \left|\frac{e^{-ih\omega}-1}{h}\omega^n\right|\leq|\omega|^{n+1}
 \end{eqnarray}
and $|\mu|_n$ is finite for $n=0,1,\ldots$ , we were able to use Lebesgue theorem in (\ref{Bu}).
For $h\in{\Bbb R}$ we have the estimate
 \begin{eqnarray}
\lefteqn{ \left|\widehat{\sigma}(t+h)-\sum\limits_{k=0}^{n-1}\frac{h^k}{k!}\,\frac{d^k}{dt^k}\,
 \widehat{\sigma}(t)\right|} \nonumber\\
  &&\displaystyle{=\left|\widehat{\sigma}(t+h)-\int\sum\limits_{k=0}^{n-1}
 \frac{(-ih\omega)^k}{k!}\,e^{-it\omega}\,
 d\sigma(\omega)\right|}\nonumber\\
  &&\displaystyle{=\left|\int\left(e^{-i(t+h)\omega}-\sum\limits_{k=0}^{n-1}\,
 e^{-it\omega}\,\frac{(-ih)^k}{k!}\,
 \omega^k\right)\,d\sigma(\omega)\right|}\nonumber\\
 &&\leq\frac{1}{n!}\int|h\omega|^n\,d\sigma(\omega)=
 \frac{|h|^n}{n!}|\mu|_n,\label{Bv}
 \end{eqnarray}
where for the last inequality we used
 \begin{eqnarray}\label{Bx}
&\left|e^{-ih}-\sum\limits_{k=0}^{n-1}\,\frac{(-i h)^k}{k!}\right|\leq\ \frac{|h|^n}{n!}\,.
 \end{eqnarray}

By the Cauchy criterion and Stirling formula
 \begin{eqnarray}\label{By}
 n!=\sqrt{2\pi n}\,\; n^n\,e^{-n}\,e^{\Theta(n)},
 \end{eqnarray}
where  $\Theta(n)<\frac{1}{12\;n}$ ,  the series $ \sum\limits_{n=0}^\infty\frac{|h|^n}{n!}|\mu|_n
$ is convergent for $|h|<R$. This and (\ref{Bv}) imply that Taylor expansion
 \begin{eqnarray}\label{Baa}
 \widehat{\sigma}(t+z)=\sum_{k=0}^\infty\frac{z^k}{k!}\;\frac{d^n}{dt^n}\widehat{\sigma}(t)
 \end{eqnarray}
is convergent for $|z|<R$. We have proved the analyticity of $\widehat{\sigma}$ on the strip
$|Im\,z|<R$. So there exist a nonempty,  maximal strip $ \{z\in{\Bbb C}:2r<Im\,z<2s\} $, such that
the characteristic function $\widehat{\sigma}(t)$ can be holomorphically extended to it. We have
shown, moreover that this strip contains the real axis i.e. $-\infty\leq2r<0<2s\leq +\infty.$ QED

Let us consider a "half" of the strip $ \{z\in{\Bbb C}:2r<Im\,z<2s\} $ i.e.
\begin{equation}\label{Bsss}
\Sigma:=\{z\in{\Bbb C}:r<Im\,z<s\}.
\end{equation}
As a consequence of the Lemma \ref{BA} we can formulate the following
\begin{proposition}\label{BB}
Under the assumptions of Lemma \ref{BA}  the map
\begin{equation}\label{Bs}
\widetilde{K}:\Sigma \ni z \longmapsto e^{-iz\;\cdot}\in L^2({\Bbb R},d\sigma)
\end{equation}
is holomorphic and its image $\widetilde{K}\left(\Sigma\right)$ is linearly dense in $L^2({\Bbb
R},d\sigma)$.
\end{proposition}

{\it Proof} :

In order to see  that the function $e^{-iz\cdot}\;$ belongs to $L^2({\Bbb R},d\sigma)$ let us
notice that
 \begin{eqnarray}\label{Bab}
\int  \left|e^{-iz\omega}\right|^2\,d\sigma(\omega)&=&\int
e^{-i(z-\bar{z})\omega}\,d\sigma(\omega)\nonumber\\
&=&\widehat{\sigma}(z-\bar{z})\,<+\infty
 \end{eqnarray}
for $(z-\bar{z})\in \{z\in{\Bbb C}:2r<Im\,z<2s\} $. Thus in the basis $\{P_n\}_{n=0}^\infty$ we
have
 \begin{eqnarray}\label{Bac}
\widetilde{K}(z)=e^{-iz\cdot}=\sum_{n=0}^\infty\widehat{\sigma}_n(z)P_n(\cdot)\,,
 \end{eqnarray}
where the coefficients functions $\widehat{\sigma}_n$ are holomorphic extensions of
\begin{eqnarray}\label{Bm}
\widehat{\sigma}_n(t):=\int e^{-it\omega}P_n(\omega)\,d\sigma(\omega)
\end{eqnarray}
onto the whole strip $\Sigma$. Thus the map $ \widetilde{K}$ is complex analytic map of the strip
$\Sigma$ into Hilbert space $L^2({\Bbb R}, d\sigma)$.

In order to show that $\widetilde{K}(\Sigma)$ is linearly dense in $L^2({\Bbb R},d\sigma)$ let us
notice that the monomials
 \begin{eqnarray}\label{Bad}
\omega^n=i^n\;\frac{d^n}{dz^n}\,\widetilde{K}(z)(\omega)\mid _{z=0},
 \end{eqnarray}
where $n=0,1,\ldots\;\;$, belong to the linear closure of $\widetilde{K}(\Sigma)$. Since, they form
linearly dense subset of $L^2({\Bbb R},d\sigma)$ and  the same property is shared by
$\widetilde{K}(\Sigma). \;\;\;\;\;$QED
\newline

Combining (\ref{Bb}) with (\ref{Bac}) we obtain a holomorphic map
 \begin{eqnarray}\label{Bae}
\lefteqn{ \!\!\!K:=\!\mathbf U^{-1}\!\circ\!\widetilde{K}\!:\Sigma\ni z \longmapsto
|z\rangle\!:=\!\sum_{n=0}^\infty\! \widehat{\sigma}_n(z)|n\rangle\!\in \!\cal{F}}
 \end{eqnarray}
of  $\Sigma$ into  Fock space $\cal{F}$. Following \cite{Odz 1}, \cite{Odz 2} we shall call
$K:\Sigma\rightarrow\cal{F}$ the coherent states map related to the quantum system described by the
Hamiltonian $\mathbf H_I$. The states $|z\rangle$, where $z\in \Sigma$, will be called
\textbf{spectral coherent states}. The coherent states map has nice physical properties and, as we
will show later, it is useful for the calculations of physical characteristics of the system.

By the formulae
 \begin{eqnarray}\label{Baf}
 \Omega&:=&i\frac{\partial^2}{\partial\bar{z}\partial
 z}\left(\;\log\widehat{\sigma}(z-\bar{z})\;\right)\,d\bar{z}\wedge d
 z\nonumber\\
 &=&-\frac{1}{2}\,\frac{d^2}{dy^2}
 \left(\;\log\widehat{\sigma}(2iy)\;\right)\,dx\wedge dy,
 \end{eqnarray}
where $z=x+iy$, we will define the symplectic form $\Omega$ on $\Sigma$. Using the mean value
function
 \begin{eqnarray}\label{Bag}
 \langle\mathbf  H_I\rangle_z:=\frac{\langle z|\mathbf
 H_I|z\rangle}{\langle
 z|z\rangle}=-\frac{1}{2}\;\frac{d}{dy}\,\log\widehat{\sigma}(2iy)
 \end{eqnarray}
of the Hamiltonian  in spectral coherent states $| z\rangle\,,z\in\Sigma$, we define the classical
Hamiltonian system
 \begin{eqnarray}\label{Bah}
 X_{\langle \mathbf H_I\rangle_z}\lrcorner\;
\Omega=d\langle\mathbf H_I\rangle_z
 \end{eqnarray}
on the symplectic manifold $(\Sigma,\Omega)$. The Hamiltonian flow, tangent to the vector field $
X_{\langle \mathbf H_I\rangle_z}$ is given by (\ref{Br}). Let us denote by ${\Bbb CP}(\cal{F})$ the
complex projective Hilbert space modelled on the Fock space $\cal{F}$. Let $\Omega_{FS}$ denote
Fubini-Study $(1,1)-$form on ${\Bbb CP}({\cal{F}})$,(for the definition of $\Omega_{FS}$ consult
for example  \cite{G-H}). The form $\Omega_{FS}$ is closed and nonsingular. So, $({\Bbb
CP}({\cal{F}}),\Omega_{FS})$ can be considered as a symplectic manifold which can be interpreted as
the quantum phase space of the system described by the Hamiltonian $\mathbf H_I$.
\begin{proposition}\label{BC}

The projectivization ${\cal K}:\Sigma\rightarrow{\Bbb CP}({\cal{F}})$, ${\cal K}(z):={\Bbb
C}|z\rangle\,$ of the coherent states map (\ref{Bae}) is the holomorphic symplectic map, i.e.
 \begin{eqnarray}\label{Bai}
{\cal K}\,^*\Omega_{FS}=\Omega
 \end{eqnarray}
and the diagram
\begin{equation}\label{Baj}
\begin{array}{ccc}
&\Sigma  \stackrel{{\cal K}}{\longrightarrow}{\Bbb CP}({\cal F}) &  \\
&T_t\uparrow\;\;\;\;\;\;\;\;\;\;\;\;\uparrow e^{-i\mathbf  H_It}&\\
&\Sigma  \stackrel{{\cal K}}{\longrightarrow}{\Bbb CP}({\cal F}) &  \\
\end{array}
\end{equation}
is commutative for any $t\in{\Bbb R}$.
\end{proposition}
{\it Proof}
 \newline The equality (\ref{Bai}) can be checked by direct
calculation. The commutativity of the diagram (\ref{Baj}) follows from (\ref{Bac}) by the  use of
the formulae for quantum (\ref{Bj}) and classical (\ref{Br}) evolution of the system. QED

Recapitulating: we see that the coherent states map maps symplectically the classical phase space
$\Sigma$ of the system $(\Sigma,\Omega,\langle \mathbf H_I\rangle_z)$ into the quantum phase space
${\Bbb CP}({\cal F})$ of the system $\left({\Bbb CP}({\cal F}),\Omega_{FS},\mathbf H_I\right)$. It
is equivariant with respect to the classical and the quantum flows. The mean value function
$\langle \mathbf H_I\rangle_z$ of the quantum Hamiltonian $\mathbf H_I$ give the classical
Hamiltonian of the system. So, the above picture is analogous to the one related to the harmonic
oscillator (see \cite{Sch}). For the general theory of quantization and description of  physical
systems in terms of the coherent states map see \cite{Odz 1}. The model of the physical system
considered here gives an important and interesting example illustrating the theory which was
developed in \cite{Odz 2}.

Let us define \textbf{spectral annihilation operator} $\pmb\alpha$ by the condition
\begin{equation}\label{Bak}
\pmb \alpha| z\rangle=z|z\rangle,
\end{equation}
which means that $\pmb\alpha$ has the spectral coherent states $|z\rangle$ as eigenvectors with
eigenvalues $z\in\Sigma$. It is defined on the dense linear domain, spanned by  spectral coherent
states. The representation in $L^2({\Bbb R},d\sigma)$ is given by
\begin{equation}\label{Bal}
\left(\mathbf U\circ\pmb\alpha\circ\mathbf U^{-1}\psi\right)(\omega)\equiv
i\frac{d}{d\omega}\psi(\omega).
\end{equation}
The domain $D\left(\mathbf U\circ\pmb\alpha\circ\mathbf U^{-1}\right)$ is given as a vector space
of all polynomials.

According to \textbf{Proposition \ref{BA}},   the spectral coherent states form a linearly dense
subset in ${\cal F}$. Hence one can define anti-linear monomorphism $\overline{\mathbf U}$
\begin{equation}\label{Bam}
{\cal F}\ni|\psi\rangle\longmapsto\overline{\mathbf U}|\psi\rangle:=\langle\psi|K(\cdot)\rangle
\in{\cal O}(\Sigma)
\end{equation}
of the Fock space $\cal{F}$ into vector space ${\cal O}(\Sigma)$ of holomorphic functions on
$\Sigma$. In such a way we obtain the third realization of Hilbert space of states, this time as
the space of holomorphic functions $\overline{\mathbf U}({\cal F})\subset{\cal O}(\Sigma)$ with the
scalar product defined by
\begin{equation}\label{Ban}
\langle\Phi|\Psi\rangle\equiv\langle\overline{\mathbf U}\left(|\phi\rangle\right),\overline{\mathbf
U}\left(|\psi\rangle\right)\rangle:= \langle\psi|\phi\rangle,
\end{equation}
where  $\Psi=\overline{\mathbf U}|\psi\rangle,\;\Phi=\overline{\mathbf U}|\phi\rangle$.
\begin{proposition}\label{BD}
Let the measure
 \begin{equation}\label{Bao}
 d\mu(\bar{z},z)=\mu(y)dxdy,
 \end{equation}
on $\Sigma$, ($z=x+iy$) be such that the weight function $\mu$
 satisfies
\begin{equation}\label{Bap}
\frac{d\sigma}{d\omega}(\omega)\int_r^s dy\,\mu(y)e^{2y\omega}=1
\end{equation}
for $\omega\in supp\; d\sigma$.

 Then the scalar product (\ref{Ban}) can be expressed by the
integral
\begin{equation}\label{Bar}
\langle\psi|\phi\rangle=\int_\Sigma \overline{\Psi}(z)\Phi(z) \,d\mu(\bar{z},z).
\end{equation}
Moreover, the kernel function
\begin{equation}\label{Bas}
\langle z|v\rangle=\widehat{\sigma} (v-\bar{z})
\end{equation}
is a reproducing kernel function with respect to the measure (\ref{Bao}), i.e.
\begin{equation}\label{Bat}
\Psi(v)=\int_{\Sigma}\widehat{\sigma} (v-\bar{z})\Psi(z)\,d\mu(\bar{z},z)
\end{equation}
for any $\Psi\in \overline{\mathbf U}({\cal F})$.
 \end{proposition}
{\it Proof} :\newline
 In order to prove that (\ref{Bar}) and (\ref{Bat}) are valid for
$d\mu(\bar{z},z)$ given by (\ref{Bao}-\ref{Bap}) let us observe that $d\mu(\bar{z},z)$ has the form
(\ref{Bao}) since the kernel $\widehat{\sigma}(\cdot-\bar{z})$ is invariant with respect to the
one-parameter group of translation (\ref{Br}). Hence we have
\begin{widetext}\begin{eqnarray}
\displaystyle{\int_\Sigma
\widehat{\sigma}(v-\bar{z})\widehat{\sigma}(z-\bar{w})\,d\mu(\bar{z},z)}&=& \displaystyle{\int
d\sigma(\omega)\int d\sigma(\omega')\int_{-\infty}^\infty dx\int_r^s
\mu(y)dy\;e^{-iv\omega+i\bar{w}\omega'}e^{-ix(\omega'-\omega)}e^{y(\omega+\omega')}}\nonumber\\
&=& \displaystyle{\int d\sigma(\omega)
 \int d\sigma(\omega')\delta(\omega-\omega') \int_r^s
dy\mu(y)\;e^{y(\omega+\omega')}e^{-iv\omega+i\bar{w}\omega'}}\nonumber\\
&=&\displaystyle{\int d\sigma(\omega)
 \int\frac{d\sigma}{d\omega}(\omega+\tau)\delta(\tau)\;d\tau
 \int_r^s
dy\mu(y)\;e^{y(2\omega+\tau)}e^{-i(v-\bar{w})\omega}e^{i\bar{w}\tau}}\nonumber\\
&=& \displaystyle{\int d\sigma(\omega)\frac {d\sigma}{d\omega} (\omega) \int_r^s
dy\mu(y)\;e^{2y\omega}e^{-i(v-\bar{w})\omega}}. \label{Bau}
\end{eqnarray}\end{widetext}
If $\mu$ satisfies (\ref{Bap}) then (\ref{Bau}) takes the form
\begin{eqnarray}\label{Baw}
\int\widehat{\sigma}(v-\bar{z})\widehat{\sigma}(z-\bar{w})\,d\mu(\bar{z},z)
=\widehat{\sigma}(v-\bar{w}),
\end{eqnarray}
which the reproducing property. $\;\;\;\;$QED

In the sequel, let us assume that $\overline{\mathbf U}({\cal F})=L^2{\cal O}(\Sigma, d\mu)$, where
$L^2{\cal O}(\Sigma, d\mu)$ denotes the Hilbert space of holomorphic functions which are
square-integrable with respect to $d\mu$ on $\Sigma$. Due to this assumption $\overline{\mathbf U}$
is an anti-unitary map and the holomorphic functions
\begin{eqnarray}\label{Bav}
\overline{\mathbf U}|n\rangle=\langle n|z\rangle=\widehat{\sigma}_n(z),\;\;\;n=0,1,\ldots
\end{eqnarray}
form an orthonormal basis in $L^2{\cal O}(\Sigma, d\mu)$.

One has the commutative diagram
\begin{equation}\label{Baz}
\begin{array}{ccc}
& {\cal F} &  \\
&\mathbf U\swarrow \quad\quad\quad\searrow \overline{\mathbf U} &\\
&L^2({\Bbb R}, d\sigma)\stackrel{\overline{\mathbf U}\circ \mathbf U^{-1}}
{\longrightarrow}L^2{\cal
O}(\Sigma, d\mu)
\end{array}
\end{equation}
where the  anti-unitary map $\overline{\mathbf U}\circ \mathbf U^{-1}$ is given by
\begin{equation}\label{Bba}
(\overline{\mathbf U}\circ \mathbf U^{-1}\psi)(z)=\int
e^{-iz\omega}\overline{\psi}(\omega)\,d\sigma(\omega),
\end{equation}
where $\psi\in L^2({\Bbb R}, d\sigma)$. Thus the Hamiltonian is given by
\begin{equation}\label{Bbb}
(\overline{\mathbf U}\circ \mathbf H_I\circ \overline{\mathbf U}^{-1}\Psi)(z)\equiv
i\,\frac{d}{dz}\Psi (z)
\end{equation}
and is defined on the domain $D(\overline{\mathbf U}\circ \mathbf H_I\circ\overline{\mathbf
U}^{-1})=\left\{\Psi\in L^2{\cal O }(\Sigma,d\mu):\;\frac{d}{dz}\Psi\in L^2{\cal O
}(\Sigma,d\mu)\right\}$.

In terms of the Hilbert space $L^2{\cal O}(\Sigma, d\mu)$  it is possible to find an explicit form
of the creation operator $\pmb\alpha^*$, i.e. Hermitian conjugate of the spectral annihilation
operator $\pmb\alpha$ defined by (\ref{Bak}). We will call $\pmb \alpha^*$ the \textbf{spectral
creation operator}. Using (\ref{Bak}) and (\ref{Bam}), we obtain
\begin{eqnarray}\label{Bax}
(\overline{\mathbf U}\circ\pmb\alpha^*\circ\overline{\mathbf U}^{-1}\Psi)(z)\equiv z\Psi(z).
\end{eqnarray}
Thus we see that the domain of $\overline{\mathbf U}\circ\pmb\alpha^*\circ\overline{\mathbf
U}^{-1}$ is given by
\begin{eqnarray}
\lefteqn{D(\overline{\mathbf U}\circ\pmb\alpha^*\circ\overline{\mathbf
U}^{-1})}\nonumber\\
&&=\left\{\Psi\in L^2{\cal O}(\Sigma, d\mu):z\Psi\in L^2{\cal O}(\Sigma, d\mu)\right\}.\label{Bay}
\end{eqnarray}

Taking into the account the  above considerations let us notice that the  operator $\pmb\alpha^*$
is described explicitly in the $L^2{\cal O }(\Sigma,d\mu)$-- realization and the operator
$\pmb\alpha$ is explicitly given in $L^2({\Bbb R},d\sigma)$-- realization. They satisfy the
canonical commutation relations
\begin{equation}\label{Bbc}
\left[\mathbf H_I,\pmb\alpha\right]=\left[\mathbf H_I,\pmb\alpha^*\right]=i
\end{equation}
with the Hamiltonian $\mathbf H_I$, giving
\begin{equation}\label{Bbd}
\left[\mathbf H_I,\pmb\alpha-\pmb\alpha^*\right]=0
\end{equation}
i.e. the operator $\pmb\alpha-\pmb\alpha^*$ is an integral of motion for the system under
consideration.

From a physical point of view (see (\ref{1}, \ref{2}, \ref{3})) it is important to describe the
time evolution in interaction picture of the system, i.e. $e^{-i\mathbf H_It}|\psi(0)\rangle$.  To
do this let us introduce the following notation for the matrix elements of $e^{-i\mathbf H_It}$
\begin{equation}\label{Bbe}
\widehat{\sigma}_{m,n}(t):=\left\langle m| e^{-i\mathbf H_I t}
 n\right\rangle\!=\!\int
e^{-i\omega t }P_m(\omega)P_n(\omega)\,d\sigma(\omega).
\end{equation}
Note that
\begin{eqnarray}\label{Bbf}
\widehat{\sigma}_{m,n}(t)=P_m(i\,\frac{d}{dt})\widehat{\sigma}_n(t),
\end{eqnarray}
where $\widehat{\sigma}_n(t)$ are given by (\ref{Bm}) and they satisfy
\begin{eqnarray}\label{Bbg}
\widehat{\sigma}_n(t)=\frac{1}{P_0}\left\langle n|e^{-i\mathbf H_I t
}0\right\rangle=P_n(i\,\frac{d}{dt})\widehat{\sigma}(t).
\end{eqnarray}
The interaction evolution in the space $\cal F$ is thus given by
\begin{eqnarray}\label{Bbh}
e^{-i\mathbf H_It}|\psi\rangle=\sum\limits_{m,n=0}^\infty\langle
m|\psi\rangle\widehat{\sigma}_{m,n}(t)|n\rangle,
\end{eqnarray}
while in the space $L^2({\Bbb R}, d\sigma)$ the evolution is described by (\ref{Bj}). In $L^2{\cal
O}(\Sigma, d\mu)$-realization we have
\begin{equation}\label{Bbi}
\left(\overline{\mathbf U}\circ e^{-i\mathbf H_It}\circ \overline{\mathbf
U}^{-1}\right)\Psi(z)=\Psi(z+t).
\end{equation}
As a consequence of (\ref{Bbe}) and  (\ref{Bbi}) we obtain the relation
  \begin{eqnarray}\label{Bbii}
\widehat{\sigma}_{m,n}(z_1+z_2)=\sum_k\widehat{\sigma}_{m,k}(z_1)\widehat{\sigma}_{k,n}(z_2),
  \end{eqnarray}
 which for $m=0$  can be expressed in the form
 \begin{eqnarray}\label{Bbjj}
\widehat{\sigma}_n(z_1+z_2)=\sum_k\widehat{\sigma}_{k,n}(z_1)\widehat{\sigma}_k(z_2).
 \end{eqnarray}
Moreover, putting  $m=n=0$ in (\ref{Bbii}) we obtain the formulae (\ref{Bas}) for the reproducing
kernel.

At the end of Section 3 it was  shown (see(\ref{1}), (\ref{2}), (\ref{3})) that the quantities
 \begin{eqnarray}\label{4}
 \langle\psi(0)|\,e^{i\mathbf H_It}\,\mathbf F\,e^{-i\mathbf H_It}
\psi(0)\rangle
  \end{eqnarray}
plays an important role if we consider the expectation values of the operator $\mathbf F$ on the
time evolving state $|\psi(t)\rangle$ (see (\ref{Aaf})).

 The variety of the realizations of our model, namely, the
${\cal{F}},\;L^2({\Bbb R},d\sigma)$ and $L^2{\cal O}(\Sigma, d\mu)$ representations allow us to
give three equivalent formulae on (\ref{4})
 \begin{eqnarray}
\lefteqn{\langle\psi(0)|e^{i\mathbf H_It}\mathbf{F}e^{-i\mathbf
H_It}\,\psi(0)\rangle\label{Bbj}}\\
&&=\sum\limits_{m,n,k,l}\langle\psi(0)|m\rangle\widehat
 {\sigma}_{m,n}^{\,*}(t)\langle n|\mathbf F\,k\rangle\widehat
 {\sigma}_{k,l}(t)\langle l|\psi(0)\rangle\nonumber\\
 &&= \displaystyle{\int e^{-i\omega
 t}\overline{\psi(\omega)}\left(\mathbf U\circ \mathbf F\circ
 \mathbf U^{-1}\right)\left(e^{i\omega t}\psi(\omega)\right)d\sigma(\omega)}\nonumber\\
 &&=\int_\Sigma\overline{\Psi(z+t)}\left(\overline{\mathbf U}\circ
 \mathbf F\circ \overline{\mathbf
 U}^{\,-1}\right)\left(\Psi(z+t))d\mu(z,\overline{z}\right),\nonumber
 \end{eqnarray}
where $\psi=\mathbf U|\psi(0)\rangle$ and $\Psi=\overline{\mathbf U}|\psi(0)\rangle$. In this way
we have a very strong instrument for calculations of many physical characteristics of the system
under consideration.

 In particular we have
  \begin{equation}
e^{i\mathbf H_It}\pmb\alpha e^{-i\mathbf H_It}=\pmb\alpha+t\label{Bbjjj}
  \end{equation}
  and therefore
\begin{equation}
\langle\psi(0)|e^{i\mathbf H_It}\pmb\alpha e^{-i\mathbf
H_It}\,\psi(0)\rangle=\langle\psi(0)|\pmb\alpha\psi(0)\rangle+t\langle\psi(0)|\psi(0)\rangle.
  \end{equation}
\section{Integrable systems related to classical orthogonal polynomials}
\renewcommand{\theequation}{5.\arabic{equation}}
\setcounter{equation}{0}

 Here we shall investigate the classes of the physical systems
with Hamiltonians of the form (\ref{Abk}) with the coefficients $b(n)$ and $h(n)$ given in table
\ref{tab:tableA.3.}. The three classes of Hamiltonian operators are related to Hermite, Laguerre
and Jacobi polynomials. We choose one mode case for simplicity and the circumstances which make the
reduction not necessary. Then the Hamiltonians are expressed in terms of usual creation and
annihilation operators in the following form:
\begin{widetext}\begin{subequations}\label{ca}\addtocounter{equation}{0}
 \begin{eqnarray}
\mathbf{H}_I^{Her}&:=&-\frac{a_0}{a_1}+\sqrt{-\frac{b_0}{a_1}}\;(\mathbf{a^{}}+\mathbf{a^*}),
\label{Ca}
 \end{eqnarray}
 \begin{eqnarray}
\mathbf{H}_I^{Lag}&:=&-\frac{b_1}{a_1}\mu-\frac{b_0}{b_1}-\frac{2b_1}{a_1}\;\mathbf{a^*a^{}}-
\frac{b_1}{a_1}\sqrt{\mathbf{a^*a^{}}+\mu}\;\;\mathbf{a^{}}
-\frac{b_1}{a_1}\sqrt{\mathbf{a^*a^{}}+\mu+1}\;\;\mathbf{a^*}\label{Cb},
 \end{eqnarray}
 \begin{eqnarray}
\mathbf{H}_I^{Jac}&:=&\frac{2\mathbf{a^*a^{}}(a+b)(\mu +\nu-1)+2(\mathbf{a^*a^{}})^2(a+b)
   -2b\mu-2a\nu+\mu \nu (a+b)+b\mu^2+a\nu^2}
  {(\mu+\nu-2+2\mathbf{a^*a^{}})(\mu+\nu+2\mathbf{a^*a^{}})}\nonumber\\
&& +(b-a)\sqrt{\frac{ (\mu+\mathbf{a^*a^{}})(\nu+\mathbf{a^*a^{}}) (\mu+\nu+\mathbf{a^*a^{}}-1)}
{(\mu+\nu+2\mathbf{a^*a^{}}-1)(\mu+\nu+2\mathbf{a^*a^{}})^2(\mu+\nu+2\mathbf{a^*a^{}}+1)}}
\;\;\mathbf{a^{}}\nonumber\\
&&+(b-a)\sqrt{\frac{(\mu+\mathbf{a^*a^{}}+1)(\nu+\mathbf{a^*a^{}}+1)
  (\mu+\nu+\mathbf{a^*a^{}})}
{(\mu+\nu+2\mathbf{a^*a^{}}+1)(\mu+\nu+2\mathbf{a^*a^{}}+2)^2(\mu+\nu+2\mathbf{a^*a^{}}+3)}}
\;\;\mathbf{a^*}.\label{Cc}
 \end{eqnarray}
\end{subequations}\end{widetext}
The ranges of the parameters $\mu,\;\nu,\;b_0,\;$ and $a_1$ are chosen such that the operators are
well defined and are essentially selfadjoint. In $L^2({\Bbb R},d\sigma)$ (i.e. spectral)
representation the formulae \eqref{Abk} lead to three-therm recurrence relation (\ref{Be}) (see
also (\ref{A.10})).

From Pearson equation (see (\ref{A.4}) and table \ref{tab:tableA.1.}) we obtain the expressions for
measures:
\begin{subequations}\label{cb}
\begin{eqnarray}
d\sigma^{Her}(\omega)=C e^{\frac{a_1}{2b_0}(\omega+\frac{a_0}{a_1})^2}\;d\omega\label{Cd}
\end{eqnarray}
for $\omega\in {\Bbb R}$,
\begin{eqnarray}
d\sigma^{Lag}(\omega)=C\left(\omega+\frac{b_0}{b_1}\right)^{\mu-1}
e^{\frac{a_1}{b_1}\omega}\;d\omega\label{Ce}
\end{eqnarray}
for $\omega\in \left(-\frac{b_0}{b_1},\infty\right)$,
\begin{eqnarray}
d\sigma^{Jac}(\omega)=C(\omega-a)^{\mu-1}(b-\omega)^{\nu-1}\;d\omega\label{Cf}
\end{eqnarray}
for $\omega\in \left(a,b\right) .$
\end{subequations}

In the holomorphic representation $L^2{\cal O}(\Sigma, d\mu)$ all Hamiltonians act as derivations:
${i\frac{d}{dz}}$, (see formulae (\ref{Bbb})) but the difference between the systems is hidden in
the reproducing measures $d\mathbf{\mu}(\bar{z},z)=\mathbf{\mu}(y)\,dxdy,\;(z=x+iy)$, and the
choice of the domain $\Sigma$. The general case is described in Proposition \ref{BD}. Here we solve
equation (\ref{Bap}) for $\mathbf{\mu}(y)$ in  the special class, namely: continuous functions
except, possibly finite number of points in every compact subset. The discontinuity points are
assumed to be of first kind. Let us summarize the results in the following:\newline
\begin{enumerate}
  \item [\textbf{H)}] \textbf{ Hermite case}:  $\Sigma={\Bbb C}$ and
 $$\addtocounter{equation}{1}
\mathbf{\mu}^{Her}(y)=\frac{1}{C}\;e^{-\frac{a_0^2}{2b_0a_1}}\sqrt{-\frac{a_1}{2b_0\pi}}
e^{\frac{2b_0}{a_1}(y+\frac{a_0}{2b_0})^2},\label{Cg}\eqno(\theequation.a)
$$
 \textbf{ \item [L)] Laguerre case}:\newline $\Sigma=
\left\{z=x+i\,y\in{\Bbb C}:y<-\frac{a_1}{2b_1}\right\}$ and for $\mu>1$
$$\label{Ch}
\mathbf{\mu}^{Lag}(y)=\frac{2\;e^{\frac{b_0a_1}{b_1^2}}}{C\,\Gamma(\mu-1)}
\left(-2y-\frac{a_1}{b_1}\right)^{\mu-2}e^{\frac{2b_0}{b_1}y}.\eqno(\theequation.b)
$$
\end{enumerate}
For $\mu=1$ we obtain an isomorphism of $L^2{\cal O}(\Sigma,d\mu)$ with $H^2(D,d\lambda)$ - the
Hardy class of functions on the unit disc $D\subset\Bbb C$ with the measure $d\lambda$ supported on
the circle $\partial D= \{e^{i\varphi}:\varphi\in[0,2\pi]\}$ and given by
\begin{equation}
d\lambda=\frac{1}{1-\sin \varphi}d\varphi.\label{Cj}
\end{equation}
\begin{enumerate}
 \item [\textbf{  J)}] \textbf{Jacobi case}: $\Sigma={\Bbb C}$ and for
$\mu+\nu>3$\begin{widetext}
 \addtocounter{equation}{-2}
 \begin{subequations}\addtocounter{equation}{2}
\begin{eqnarray}
\mathbf{\mu}^{Jac}(y)&=&\left\{
\begin{array}{l} \displaystyle{
    \frac{2}{C}\frac{(b-a)^{1-\frac{\mu+\nu}{2}}}
    {\Gamma(\mu-1)}e^{-(b+a)y}(2y)^{\frac{\mu+\nu}{2}-2}\,
    W_{\frac{\nu-\mu}{2},\frac{3}{2}-
    \frac{\mu+\nu}{2}}[2(b-a)y]\quad\quad\;\;\;\;\textrm{for}\;\;y>0}\\
     \displaystyle{\frac{2}{C}\frac{(b-a)^{1-\frac{\mu+\nu}{2}}}
    {\Gamma(\nu-1)}e^{-(b+a)y}(-2y)^{\frac{\mu+\nu}{2}-2}\,
    W_{\frac{\mu-\nu}{2},\frac{3}{2}-
    \frac{\mu+\nu}{2}}[-2(b-a)y]\quad\;\;\textrm{for}\;\;y<0}
   \end{array}\right.
\end{eqnarray}
\end{subequations}\end{widetext}
\end{enumerate}
where $W_{\kappa,\lambda}(z)$ are confluent hypergeometric Whittaker's functions (for definition
see \cite{A-S}). This formula simplifies in the case $\mu=\nu$ corresponding to the Gegenbauer
polynomials. The following statement is true for a larger domain of parameter $\mu$, namely for
$\mu>1$,
 \addtocounter{equation}{+1}
\begin{eqnarray}
\lefteqn{\mathbf{\mu}^{Geg}(y)=\frac{2}{C}\left(\frac{-2y}{b-a}\right)^{\mu-\frac{3}{2}}}\nonumber\\
&&\;\;\;\;\;\;\;\;\;\;\times\;
e^{-(b+a)y}\frac{1}{\Gamma(\mu-1)\sqrt{\pi}}K_{\mu-\frac{3}{2}}\left((a-b)y\right)
\end{eqnarray}
with $K_\alpha(z)$ being the modified Bessel functions (for definition see \cite{A-S}).

 For all three cases one can find the explicit form of matrix
elements of propagator (\ref{Bbe}). Because of the relations (\ref{Bbg}) and (\ref{Bbf}), we should
display the characteristic functions  (\ref{Bk}) first:
\begin{subequations}\label{CCC}
 \begin{eqnarray}
\widehat{\sigma}^{Her}(z)= C\,\sqrt{-\pi \frac{ 2b_0}{a_1}}\,e^{\,\frac{a_0^2}{2a_1b_0}} \,
e^{\,\frac{b_0}{2a_1}(z+i\frac{a_0}{b_0})^2}, \label{Ck}
 \end{eqnarray}
  \begin{eqnarray}
\widehat{\sigma}^{Lag}(z)=C\;\Gamma(\mu)\,
  e^{-\frac{b_0a_1}{b_1^2}}\left(-\frac{z}{i}-\frac{a_1}{b_1}
  \right)^{-\mu}e^{\frac{b_0}{b_1}i z},\label{Cl}
    \end{eqnarray}
     \begin{eqnarray}
\lefteqn{\widehat{\sigma}^{Jac}(z)\equiv\widehat{\sigma}_J(z;\mu,\nu)\nonumber=C\;\frac{\Gamma(\mu)
\Gamma(\nu)}{\Gamma(\mu+\nu)}}\\
&&\;\;\;\;\;\;\;\;\times(b-a)^{\mu+\nu-1} e^{-iaz}
 \,_1F_1(^{\;\;\mu}_{\mu+\nu};(a-b)iz)\label{C³}.
 \end{eqnarray}
\end{subequations}
  The symbols $\widehat{\sigma}^{Jac}(z;\mu,\nu)$ are introduced
in order to simplify the next formulae. Using the Rodrigues formula (see (\ref{A.7})) we obtain the
explicit form of $\widehat{\sigma}_n(z)$:
\begin{subequations}\label{cm}
 \begin{eqnarray}
\widehat{\sigma}^{Her}_{n}(z)&=&c^{Her}_n(ib_0z)^n\,\widehat{\sigma}^{Her}(z)\label{Cm},\\
\widehat{\sigma}^{Lag}_{n}(z)&=&c^{Lag}_n\!\left(\!\frac{b_1z}{z+i\frac{a_1}{b_1}}\!\right)^{\!n}\!\frac{\Gamma
(\mu+n)}{\Gamma(\mu)} \,\widehat{\sigma}^{Lag}(z)\label{Cn},\\
\widehat{\sigma}^{Jac}_{n}(z)&=&c^{Jac}_n(ib_2z)^n\,\widehat{\sigma}^{Jac}(z;\mu+n,\nu+n)\label{Co}.
 \end{eqnarray}
 \end{subequations}
After a simple but tedious calculation we find:
\begin{widetext}\begin{subequations}\label{cp}
 \begin{eqnarray}
\widehat{\sigma}^{Her}_{m,n}(z)&=&e^{\frac{b_0}{2a_1}(z+i\frac{a_0}{b_0})^2}
  e^{\frac{a^2_0}{2a_1b_0}}
  (iz)^{m+n}\sqrt{\left(-\frac{b_0}{a_1}\right)^{m+n}}\sqrt{m!n!}
\sum_{k=0}^{min\{m,n\}}
  \frac{\left(\frac{a_1} {b_0}\right)^kz^{-2k}}{(m-k)!(n-k)!k!}\label{Cp}\;,
 \end{eqnarray}
 \begin{eqnarray}
\widehat{\sigma}^{Lag}_{m,n}(z)&=&c^{Lag}_mc^{Lag}_n\,b_1^{m+n}\frac{\Gamma(\mu+m)
\Gamma(\mu+n)}{\Gamma^2(\mu)} \,\widehat{\sigma}^{Lag}(z)
\sum_{k=0}^m\left(^m_k\right)
\left(\frac{ia_1}{b_1z+ia_1}\right)^k\,_2F_1\left(^{\mu+k,-n}_{\;\;\;\;\mu}
 ;\frac{ia_1}{ia_1+b_1z}\right)\label{Cr}\;,
  \end{eqnarray}
 \begin{eqnarray}
\widehat{\sigma}^{Jac}_{m,n}(z)&=&c^{Jac}_mc^{Jac}_n\,(-b_2)^{m+n}
\sum_{k=0}^m\sum_{l=0}^n\left(^m_k\right)\left(^n_{\,l}\right)(-1)^{k+l}
\frac{\Gamma(\mu+m)\Gamma(\mu+n)}{\Gamma(\mu+m-k)\Gamma(\mu+n-l)}
\frac{\Gamma(\nu+m)\Gamma(\nu+n)}{\Gamma(\nu+k)
\Gamma(\nu+l)}\nonumber\\
&&\;\;\;\;\;\;\;\;\;\;\;\;\;\;\;\;\;\;\;\;\; \;\;\;\;\;\;\;\;\;\;\;\;\;\;\;\;\;\;\;\;\;\;\;\;\;
\times\,\widehat{\sigma}^{Jac}(z;\mu+m+n-k-l,\nu+k+l)\label{Cs}\,.
 \end{eqnarray}
\end{subequations}\end{widetext}

 The physical quantities which are of great importance are: the
Hamiltonians $\mathbf{H}$ and $\mathbf{H}_I$, the creation $\mathbf{a^*}$ and annihilation
$\mathbf{a}$ operators, and the occupation number operator $\mathbf{N=a^*a^{}}$. In our case the
operators $\mathbf{A^{}},\;\mathbf{A^*}$ are also important. They can be interpreted as the cluster
annihilation and cluster creation operators. Similarly, the operators ${\pmb\alpha}$ and
${\pmb\alpha^*}$  which are related to the spectral coherent states map (\ref{Bae}) are interested,
too. Their physical meaning is partially explained  by the commutation relations (\ref{Bbc}). They
are related to $\mathbf{H_I,\;a^{}}$ and $\mathbf{a^*}$ in the following way:
\begin{subequations}\label{ct}
 \begin{eqnarray}
\pmb{\alpha}^{Her}=\frac{-i\, a_1}{\sqrt{-a_1b_0}}\,\mathbf{a}\;,\label{Ct}
 \end{eqnarray}
\begin{eqnarray}
\lefteqn{(b_0+b_1\,\mathbf{H}^{Lag})\pmb{\alpha}^{Lag}\nonumber}\\
&&=-i\,b_1\,\mathbf{a^*a}+i\sqrt{b_1^2( \mathbf{a^*a}+\mu)}\;\,\mathbf{a}\;,\label{Cu}
 \end{eqnarray}\begin{widetext}
\begin{eqnarray}
\displaystyle{\left(a-\mathbf{H}^{Jac}\right)\left(b-\mathbf{H}^{Jac}\right)\pmb{\alpha}^{Jac}
}&&\displaystyle{=i\,\mathbf{a^*a}\frac{ b(2\mu-\nu+\,\mathbf{a^*a}-1)
+a(2\nu-\mu+\,\mathbf{a^*a}-1)}
{\mu+\nu+2\,\mathbf{a^*a}-2}-i\,\mathbf{H}^{Jac}\,\mathbf{a^*a}\nonumber}\\
&&\;\;\;\;\;\displaystyle{-i\,\frac{(b-a)(-\mu-\nu+\,\mathbf{a^*a}-1)(\mu+\,\mathbf{a^*a})(\nu+\,\mathbf{a^*a})}
{|b_2|(\mu+\nu+2\,\mathbf{a^*a}-1)^2(\mu+\nu+2\,\mathbf{a^*a})^3}\sqrt{\frac{\mu+\nu+2\,
\mathbf{a^*a}+1}{(\mu+\nu+\,\mathbf{a^*a}-1)}}\;\mathbf{a}\,.}\label{Cw}
 \end{eqnarray}\end{widetext}
\end{subequations}

In the spectral representation of $\mathbf H_I$ the operator $\pmb\alpha$ is given  for all the
cases by $i\frac{d}{d\omega}$ but the conjugates are given by different formulae:
 \begin{subequations}\label{cv}
 \begin{eqnarray}
\left({\pmb\alpha}^{Her}\right)^*\!\!&\!\!=\!\!&\!\!-i\left(\frac{a_1}{b_0}
\omega+\frac{a_0}{b_0}+\frac{d}{d\omega}\right)\;, \label{Cv}
 \end{eqnarray}
 \begin{eqnarray}
\left({\pmb\alpha}^{Lag}\right)^*=-i\left(
\frac{a_1\omega+a_0-b_1}{b_1\omega+b_0}+\frac{d}{d\omega}\right) \label{Cx}
\end{eqnarray}
for $\mu>1$,
 \begin{equation}
  \left({\pmb\alpha}^{Jac}\right)^*\!=\!-i
\left(\frac{(a_1+2b_2)\omega+a_0-b_2(a+b)}{b_2(\omega-a)
(b-\omega)}+\frac{d}{d\omega}\right)\label{Cy}
 \end{equation}
for $\mu,\,\nu>1$.
 \end{subequations}

In the holomorphic representation the operator ${\pmb\alpha^*}$ is given by (\ref{Bax}), i.e. as
the operator of multiplication
 by the argument $z$. The operators
${\pmb\alpha}^{Her},\;{\pmb\alpha}^{Lag}$ and ${\pmb\alpha}^{Jac}$ are pseudodifferential ones and
we shall not express them explicitly here.

The occupation number operators $\mathbf{N}$  defined by (\ref{Abg}) take in the spectral
representation the following form:
 \begin{subequations}\label{cz}\begin{eqnarray}
\mathbf{N}^{Her}&=&\left(\omega+\frac{a_0}{a_1}\right)\frac{d}{d\omega}+\frac{b_0}{a_1}
\frac{d^2}{d\omega^2}\;,\label{Cz}\\
\mathbf{N}^{Lag}&=&\left(\omega+\frac{a_0}{a_1}\right)\frac{d}{d\omega}+\left(\frac{b_1}{a_1}\omega+\frac{b_0}{a_1}\right)
\frac{d^2}{d\omega^2}\label{Cca}.
 \end{eqnarray}
For the Jacobi case $\mathbf{N}^{Jac}$ we are able to write down only the relation
\begin{eqnarray}
\lefteqn{\mathbf{N}^{Jac}\left(\mathbf{N}^{Jac}-\mu-\nu-1\right)=(\omega-a)(b-\omega)\frac{d^2}{d\omega^2}
}\nonumber\\
&&\;\;\;\;\;\;\;\;\;\;\;\;\;\;\;\;\;\;\;\;+\left[(-\mu-\nu)\omega+\mu b+\nu a
\right]\frac{d}{d\omega}\label{Ccb}.
\end{eqnarray}
\end{subequations}
 In the holomorphic representation $\mathbf{N}$ can be expressed
as:
 \begin{subequations}\label{ccd}\begin{eqnarray}
\mathbf{N}^{Her}&=&-\frac{b_0}{a_1}\left(z-i\frac{a_0}{b_0}\right)\,z+z\frac{d}{dz}\;,\label{Ccd}\\
\mathbf{N}^{Lag}&=&i\,\left(\frac{b_1}{a_1}\mu+\frac{b_0}
{b_1}+i\frac{b_0}{a_1}z\right)\,z\nonumber\\
&&+\left(1+\frac{b_1}{a_1}\,i\,z\right)\,z\frac{d}{dz}\;\label{Cce}.
 \end{eqnarray}
\end{subequations}

Now, we will present the expectation values on the following states, interesting from the physical
point of view:
\begin{enumerate}
\item [i)] occupation number states $|n\rangle$, $n\in{\Bbb N}\cup\{0\}$,  i.e. the
eigenstates of $\mathbf{N}$, $\mathbf{N}|n\rangle=n|n\rangle$;
\item [ii)] Gaussian coherent states $|\zeta\rangle$, $\zeta\in\Bbb
C$, i.e. the eigenstates of $\mathbf{a},\;\;\mathbf{a}|\zeta\rangle=\zeta|\zeta\rangle$;
\item [iii)] spectral coherent states $|z\rangle$, $z\in\Sigma$, i.e.  the eigenstates  of
$\pmb{\alpha}$ , $\pmb{\alpha}|z\rangle=z|z\rangle$.
\end{enumerate}
Using the operators $U$ and $\overline{U}$ one can realize these states in spectral or holomorphic
representation, too (see (\ref{Bb}), (\ref{Bs}), (\ref{Bam}), (\ref{Bav})).

Of course the Hamiltonian $\mathbf{H}_I$ does not depend on time $t$ and its mean values are given
by:
\begin{eqnarray}
&&\langle \mathbf{H}_I\rangle_n=h(n),\\
&&\langle
\mathbf{H}_I\rangle_\zeta=e^{-|\zeta|^2}\sum_{n=0}^\infty\left.\frac{|\zeta|^{2n}}{n!}\right[h(n)\nonumber\\
&&\;\;\;\;\;\;\;\;\;\;\;\;\;\;\;\;\;\;\;\;\;\;\;\;\;\;\;\;\;\;\;\;\;\;\;\;\;\;\left.
+\frac{b(n+1)}{\sqrt{n+1}}(\bar{\zeta}+\zeta)\right],\\
&&\langle \mathbf{H}_I\rangle_z=-\frac{1}{2}\frac{d}{dy}\ln\widehat{\sigma}(2iy)
,\;\;\;\;\;\;y=\frac{z-\bar{z}}{2i}.
\end{eqnarray}
The indices $n,\;\zeta,\; z$ are related to the occupation number eigenstates, Gaussian coherent
states, and spectral coherent states, respectively. The function $b(n)$ and $h(n)$ are given in
table \ref{tab:tableA.3.} and $\widehat{\sigma}(z)$ is presented in (\ref{CCC}).

The mean values of the powers of the occupation number operator are given as follows:
\begin{eqnarray}
\langle\mathbf{N}^l(t)\rangle_n&\!\!=\!\!&\sum_{k=0}^\infty |\widehat{\sigma}_{n,k}(t)|^2n^l, \\
\langle\mathbf{N}^l(t)\rangle_\zeta&\!\!=\!\!&e^{-|\zeta|^2}\sum^\infty_{m,k,n}\frac{\zeta^n\bar{\zeta}^m}
{\sqrt{m!n!}}k^l\,\overline{\widehat{\sigma}_{m,k}(t)}\widehat{\sigma}_{n,k}(t),\\
\langle \mathbf{N}^l(t)\rangle_z&\!\!=\!\!&\frac{1}{\widehat{\sigma}(z-\bar{z})}\sum_{n=0}^\infty
|\widehat{\sigma}_n(z+t)|^2n^l,
\end{eqnarray}
where $l\in\Bbb N$. It is interesting to rewrite the last formula for every polynomial class
separately\vspace{-2ex}
\begin{widetext}\vspace{-2ex}
 \begin{subequations}\begin{eqnarray}
\left\langle
\left(\mathbf{N}^{Her}(t)\right)^l\right\rangle_{z}&=&e^{\frac{b_0}{a_1}|z+t|^2}\left(-\frac{b_0}{a_1}\right)
|z+t|^2\,_{l-1}F_{l-1}\left(^{2,\ldots,2}_{1,\ldots,1};-\frac{b_0}{a_1}|z+t|^2\right),
\end{eqnarray}
\begin{eqnarray} \label{Ma}\left\langle
\left(\mathbf{N}^{Lag}(t)\right)^l\right\rangle_{z}&=&\frac{\left(-2\frac{a_1}{b_1}\frac{z-\bar{z}}{2i}+
\frac{a_1^2}{b_1^2}\right)^\mu
|z+t|^2\mu}{|z+t-i\frac{a_1}{b_1}|^{2\mu+2}}\,_lF_{l-1}\left(^{\mu+1,2\ldots,2}_{\;\;\;\;1,\ldots,1}
;\left|\frac{z+t}{z+t-i\frac{a_1}{b_1}}\right|^2\right),
\end{eqnarray}
\begin{eqnarray}
\left\langle\left(\mathbf{N}^{Jac}(t)\right)^l\right\rangle_{z}&=&\frac{1}{\widehat{\sigma}^{Jac}(z-\bar{z})}\sum_{n=0}^\infty
{c^{Jac}_n}^2|b_2(z+t)|^{2n}\,|\widehat{\sigma}^{Jac}(z+t;\mu+n,\nu+n)|^2n^l.
\end{eqnarray}
\end{subequations} We give now the formulae for the correlation
functions:
\begin{eqnarray}
 \left\langle\mathbf{a}^{*\,r}(t)\mathbf{a}^s(t)\right\rangle_n&=&\sum_{m=0}^\infty
 \overline{\widehat{\sigma}_{n,m+r}(t)}
 \widehat{\sigma}_{n,m+s}(t)\; \frac{\sqrt{
  (m+r)!(m+s)!}}{m!},
\end{eqnarray}
\begin{eqnarray}
\langle\mathbf{a}^{*\,r}(t)\mathbf{a}^s(t)\rangle_\zeta&=&e^{-|\zeta|^2}\sum_{k,m,n=0}^\infty
\frac{\bar{\zeta}^m\zeta^k}{\sqrt{m!k!}}\overline
{\widehat{\sigma}_{m,n+r}(t)}\widehat{\sigma}_{n+s,k}(t)\,\frac{\sqrt{(n+r)!(n+s)!}}{n!},
\end{eqnarray}
\begin{eqnarray}
\left\langle\mathbf{a}^{*\,r}(t)\mathbf{a}^s(t)\right\rangle_z&=&\frac{1}{\widehat{\sigma}(z-\bar{z})}
\sum_{n=0}^\infty\overline{\widehat{\sigma}_{n+r}(z+t)}\widehat{\sigma}_{n+s}(z+t)\;
\frac{\sqrt{(n+r)!(n+s)!}}{n!}.
\end{eqnarray}

Replacing the creation and annihilation operators $\mathbf{a}^*,\;\mathbf{a}^{}$ by the cluster
creation and the cluster annihilation operators $\mathbf{A}^{*}$ and $\mathbf{A}^{}$ we obtain the
functions which by analogy will be called the cluster correlation functions:
\begin{eqnarray}
\langle\mathbf{A}^{*\,r}(t){\mathbf A}^s(t)\rangle_n&=& \sum_{k,l=0}^\infty b(k+r)...
b(k+1)b(k+s)...b(k+1)\overline{\widehat{\sigma}_{n,k+r}(t)}\widehat{\sigma}_{k+s,l}(t),
\end{eqnarray}
\begin{eqnarray}
\langle{\mathbf A}^{*\,r}(t){\mathbf
A}^s(t)\rangle_\zeta&=&e^{-|\zeta|^2}\!\!\sum\limits_{k,m,l=0}^\infty\!\frac{\bar{\zeta}^m
\zeta^l}{\sqrt{m!l!}}b(k+r)... b(k+1)b(k+s)...
b(k+1)\overline{\widehat{\sigma}_{m,k+r}(t)}\widehat{\sigma}_{k+s,l}(t),
\end{eqnarray}
\begin{eqnarray} \langle
\mathbf{A}^{*\,r}(t){\mathbf
A}^l(t)\rangle_z&=&\frac{1}{\widehat{\sigma}(z-\bar{z})}\sum\limits_{k=0}^\infty
b(k+r)...b(k+1)b(k+s)... b(k+1)\overline{\widehat{\sigma}_{k+r}(z+t)}\widehat{\sigma}_{k+s}(z+t)\,.
\end{eqnarray}\end{widetext}

The time evolution of ${\pmb\alpha}$ is given by (\ref{Bbjjj}). This allows us to express the time
dependence of $\langle{\pmb\alpha}^l(t)\rangle_\psi$, (where $|\psi\rangle$ is an arbitrary state
and $l\in{\Bbb N}$), in terms of the mean values of some powers  $\pmb\alpha\equiv{\pmb\alpha}(0)$
acting on the state $|\psi\rangle$:
\begin{eqnarray}
\langle {\pmb \alpha}^l(t)\rangle_\psi=\sum_{k=0}^l\left(^{\,l}_k\right)t^k\langle
\pmb\alpha^{l-k}\rangle_\psi.
\end{eqnarray}
As a consequence we conclude that the dispersion
$\left(\bigtriangleup{\pmb\alpha}(t)\right)_\psi=\sqrt{\left\langle\psi\,|\left[\pmb\alpha^2(t)-
\langle\psi\,|\pmb\alpha(t)\psi\rangle^2 \right]\psi\right\rangle}$ of the operator
${\pmb\alpha}(t) $ in an arbitrary state $|\psi\rangle$ does not depend on time
\begin{eqnarray}
\left(\bigtriangleup{\pmb\alpha}(t)\right)_\psi\equiv\left(\bigtriangleup {\pmb\alpha}\right)_\psi.
\end{eqnarray}
The following expectations take especially simple form:
\begin{eqnarray}
\langle{\pmb\alpha}^l(t)\rangle_n&=&t^l,\\
\langle {\pmb\alpha}^l(t)\rangle_z&=&(z+t)^l.
\end{eqnarray}

Let us now see what will happen  when the intensity of electromagnetic field is sufficiently large
(the light of a strong laser). This corresponds to the limit of the large $n$ in the Hamiltonian
(\ref{Abk}) (see table \ref{tab:tableA.3.}, too). We get the following strong-field Hamiltonians
$\mathbf H_{s}$:
 \begin{subequations}
\begin{eqnarray}
\mathbf{H}^{Her}_s&\!\!=\!\!&\sqrt{-\frac{b_0}{a_1}}\left(\mathbf a^{}+\mathbf a^*\right),\\
\mathbf{H}^{Lag}_s&\!\!=\!\!&-2\frac{b_1}{a_1}\mathbf a^*\mathbf a^{}\nonumber\\
&&-\frac{b_1}{a_1}\left( \sqrt{\mathbf a^*\mathbf a^{}+1}\,\mathbf a+\sqrt{\mathbf a^*\mathbf
a^{}+2}\,\mathbf a^*\right)
,\label{Ccg}\\
\mathbf{H}^{Jac}_s&\!\!=\!\!&\frac{a+b}{2}\nonumber\\
&& +\frac{b-a}{4}\left(\!\!\frac{1}{\sqrt{\mathbf a^*\mathbf a^{}+1}}\,\mathbf a^{}+\frac{1}
{\sqrt{\mathbf a^*\mathbf a^{}+2}}\,\mathbf a^*\!\!\right)\!\!\label{Cch}.
\end{eqnarray}
\end{subequations}
These Hamiltonians belong to the respective families given by (\ref{ca}). They are obtained in the
Hermite case by putting $ a_0=0$ in (\ref{Ca}), in the Laguerre case by putting $\mu=1$ and
$b_0=-\frac{b_1^2}{a_1}$ in (\ref{Cb}), and in the Jacobi case by putting $\mu=\nu=\frac{3}{2}$ in
(\ref{Cc}).

Let us recall the definition of the phase operator $\widehat{\mathbf{\phi}}$ \cite{P-L}:
\begin{eqnarray}
&&\!\!\!\!\!\!\!\!\exp{(i\widehat{\mathbf{\phi}})}:=(\mathbf{a}^*\mathbf{a}^{}+1)^{-\frac{1}{2}}\mathbf{a},\\
&&\!\!\!\!\!\!\!\!\exp{(-i\widehat{\mathbf{\phi}})}:=\mathbf{a}^*(\mathbf{a}^*\mathbf{a}^{}+1)^{-\frac{1}{2}},\\
&&\!\!\!\!\!\!\!\!\cos{(\widehat{\mathbf{\phi}})}:=\frac{1}{2}\left(\exp{(i\widehat{\mathbf{\phi}
})}+\exp{(-i\widehat{\mathbf{\phi}})}\right).
\end{eqnarray}
We can now rewrite (\ref{Ccg}) and (\ref{Cch}):
\begin{eqnarray}
{\mathbf{H}}^{Lag}_s&\!\!=\!\!&-\frac{b_1}{a_1}\left(2\mathbf a^*\mathbf a^{}+2\mathbf a^*\mathbf
a^{}\,\cos{(\widehat{\mathbf{\phi}})}
+\exp{(i\widehat{\mathbf{\phi}})}\right)\label{Cci},\\
\mathbf{H}^{Jac}_s&\!\!=\!\!&\frac{a+b}{2}+\frac{b-a}{2}\cos{(\widehat{\mathbf{\phi}})}\label{Ccj}.
\end{eqnarray}
So, in the Jacobi case in the strong-field limit, the Hamiltonian tends, up to a constant, to the
\textbf{cosine} of the phase operator. This subcase does not depend on the choice of the ranges of
the parameters $\mu, \;\nu$.

\section{A physical remarks}
\renewcommand{\theequation}{6.\arabic{equation}}
\setcounter{equation}{0}
\subsection{Parametric modulator}\label{6a}
In order to present some physical interpretations of the Hamiltonian (\ref{Aac}) with
$\mathbf{H}_I$ given by (\ref{Aan}), let us rewrite it in the following form
\begin{widetext}\begin{eqnarray}
\mathbf H_I&=&\sum_{j=0}^M\omega_j\mathbf a^*_j\mathbf a^{} _j+h(\mathbf a_0^*\mathbf
a_0,\ldots,\mathbf a_M^*\mathbf
a_M)\nonumber\\
&&+\left( e^{it\sum\limits_{j=0}^M\omega_j\mathbf a_j^*\mathbf a_j^{}}g (\mathbf a_0^*\mathbf
a_0,\ldots,\mathbf a_M^*\mathbf a_M) \mathbf a_0^{l_0}\ldots \mathbf
a_M^{l_M}+\left[e^{it\sum\limits_{j=0}^M\omega_j\mathbf a_j^*\mathbf a_j^{}}g (\mathbf a_0^*\mathbf
a_0,\ldots,\mathbf a_M^*\mathbf a_M) \mathbf a_0^{l_0}\ldots \mathbf
a_M^{l_M}\right]^*\,\right).\label{Fa}
\end{eqnarray}\end{widetext}
The first term, which is linear in photon number operators describes the free field. The second
term, which is an arbitrary function of these operators may be treated as a generalization of the
Kerr medium description, where $\mathbf H_I=\frac{\chi}{2}\left(\left(\mathbf a^*\mathbf
a^{}\right)^2-\mathbf a^*\mathbf a^{}\right)$,\; where $\chi$ is proportional to the third-order
nonlinear susceptibility, \cite{P-L}. The terms of the  type $h(\mathbf a_0^*\mathbf
a_0,\ldots,\mathbf a_M^*\mathbf a_M)$, after the appropriate choice of the function $h$, play an
important role in the theory of the nondemolition measurment \cite{W-M}, \cite{M-W} and in the
description of many other phenomena e.g. the optical bistability effect \cite{D-W}.

The last term in (\ref{Fa}) one can interpreted as a general form of the parametric modulator
Hamiltonian. To motivate this interpretation let us recall the form of the Hamiltonian of
nondegenerate parametric amplifier \cite{W-M}, \cite{M-G 1}
\begin{eqnarray}
\label{Fb} \mathbf H&=&\omega_0\mathbf a^*_0\mathbf a^{}_0+\omega_1\mathbf a^*_1\mathbf a^{}_1\nonumber\\
&&+ig\left(e^{2i\omega t}\mathbf a_0\mathbf a_1-\left(e^{2i\omega t}\mathbf a_0\mathbf
a_1\right)^{*}\right).
\end{eqnarray}
 This Hamiltonian describes the case when the classical  pump mode
at frequency $2\omega$ interacts in a  nonlinear optical medium with two modes at frequency
$\omega_0$ and \;$\omega_1$, such that $\omega_0+\omega_1=2\omega$. If the system starts in an
initial Gaussian 2-photon coherent state $|\zeta_0\zeta_1\rangle$, the mean photon number in
$0$-mode after time $t$ is
\begin{eqnarray}\label{Fc}
\!\!\langle \mathbf a^*_0(t)\mathbf a^{}_0(t)\rangle=|\zeta_0 \cosh gt+\zeta^*_1 \sinh
gt|^2+\sinh^2 gt,
\end{eqnarray}
hence this mode is amplified. The next example is the Hamiltonian for the frequency up-converter
(\cite{W-M})
\begin{equation}\label{Fcc}
\!\!\mathbf H\!=\! \omega_0\mathbf a_0^*\mathbf a^{}_0+\omega_1\mathbf a^*_1\mathbf a^{}_1+\kappa
\left(e^{i\omega t}\mathbf a_0^*\mathbf a_1^{}+e^{-i\omega t}\mathbf a_0^{}\mathbf a_1^{*}\right),
\end{equation}
where $\omega=\omega_1-\omega_0$.

It is easy to compare (\ref{Fa}) with (\ref{Fb}) and with (\ref{Fcc}) and conclude that our
Hamiltonian is a natural generalization of that describing parametric amplification. In order to
understand that in general (\ref{Fa}) describes not only amplification but also modulation let us
notice that due to (\ref{Aap}) we can express the mean values $\langle \mathbf a^*_j(t)\mathbf
a^{}_j(t)\rangle,\;j=0,\ldots,M$ in terms of the mean values of the operators $\mathbf
A_0(t),\ldots,\mathbf A_M(t)$. But $\mathbf A_1(t),\ldots,\mathbf A_M(t)$ are the integrals of the
motion, so if our system starts at the initial state from the reduced subspace $\cal F$ (see
(\ref{Abh})) we obtain
\begin{eqnarray}\label{Fd}
\langle \mathbf a^*_j(t)\mathbf a_j^{}(t)\rangle=l_j\langle \mathbf A_0(t)\rangle+\beta_j,
\end{eqnarray}
where the constant $\beta_j$ are uniquely determined by $\lambda_1,\ldots,\lambda_M$ and the matrix
$\alpha$. This means that the mean photon number in each mode is a linear function of $\langle
\mathbf A_0(t)\rangle$ or, in other words, the strength of the light in each mode is modulated by
the function $\langle \mathbf A_0(t)\rangle$. The modulation of the  $j$-th mode depends on the
exponent $l_j$. The shape of the function $\langle \mathbf A_0(t)\rangle$ depends on the choice of
the coupling function $g (\mathbf a_0^*\mathbf a_0,\ldots,\mathbf a_M^*\mathbf a_M)$ in (\ref{Fa})
and the initial state of the system.

As an example of the modulation function $\langle \mathbf A_0(t)\rangle$  let us consider the
situation when after reduction, we obtain the case corresponding to Laguerre polynomials and the
initial state is the spectral coherent state $|z\rangle$. From (\ref{Ma}) we obtain
\begin{eqnarray}\label{Fd}
\langle \mathbf A_0(t)\rangle_z=E|z+t|^2+F,
\end{eqnarray}
where the real constants $E,F$ depend on  $\mu, \;a_1,\; b_1$ and $\lambda_{0,l}$. In this example
the modulation function is of parabolic shape. This means, that in some interval of time we have
the amplification and dumping of the light signal in others.
\subsection{Generalized squeezed states}

The special cases of the interaction evolution operators $e^{-i\mathbf{ H}_It}$ are the unitary
displacement operators \cite{K-S}
\begin{eqnarray}\label{Fd}
\mathbf D(\zeta)=\exp(\zeta \mathbf a^*-\bar{\zeta}\mathbf a),\;\;\;\zeta\in\Bbb C,
\end{eqnarray}
the unitary squeeze operators \cite{C}
\begin{eqnarray}\label{Fe}
\mathbf S(z)=\exp(\bar{z}\mathbf  a^2-z\mathbf a^{*2}),\;\;\;z\in\Bbb C
\end{eqnarray}
and the unitary two-mode squeeze operators \cite{C-S}
\begin{eqnarray}\label{Ff}
\mathbf T(\xi)=\exp(\bar{\xi} \mathbf a_0\mathbf a_1-\xi \mathbf a_0^*\mathbf
a_1^*),\;\;\;\xi\in\Bbb C.
\end{eqnarray}
This means that the Glauber coherent states and the squeezed states are special cases of the
spectral coherent states defined in Section 4. In such a way, the two concepts of the notion of the
coherent states meet each other in our framework. The first one, presented in \cite{Sch},
\cite{K-S}, \cite{C}, \cite{C-S}, is related to the minimalization of the suitable uncertainly
relations. The second one, presented in \cite{Odz 1}, \cite{Odz 2} is based on the symplectic
embedding of the classical phase space of the system into the quantum phase space (equipped with
the Fubbini-Study symplectic form).
\section*{Acknowledgement}
We would like to thank prof. J. Tolar and dr. Z. Hasiewicz for careful reading and suggestions that
made our paper better. This work was supported in part by KBN grant 2 PO3 A 012 19.
\section*{ Appendix}
\renewcommand{\theequation}{A.\arabic{equation}}
\renewcommand{\theproposition}{A.1}
\setcounter{equation}{0}

 Here we present some facts from the theory of classical
polynomials.

Let us consider a pair of real polynomials $(A(\omega),B(\omega))$ of degree not greater than one
and two, respectively
\begin{eqnarray}\label{A.1}
A(\omega)&:=&a_1\omega+a_0, \;\;\;\;\;\;\;\;\;\;\;\;\;\;\;\;\,\;\;a_i\in\Bbb R,\\
B(\omega)&:=&b_2\omega^2+b_1\omega+b_0, \;\;\;\;\;\;\;\;b_i\in\Bbb R.\label{A.1.1}
\end{eqnarray}
 The Pearson equation associated  with $(A(\omega),B(\omega))$ on
the interval $(a,b)\subset{\Bbb R}\;\; (-\infty\leq a<b\leq+\infty)$ is the differential equation
for the weight function $\varrho$:
\begin{equation}\label{A.2}
\frac{d}{d\omega}\left(\varrho B\right)=\varrho A
\end{equation}
with the boundary conditions
\begin{equation}\label{A.3}
\varrho(a) B(a)=0=\varrho(b) B(b).
\end{equation}
Each family of classical orthogonal polynomials $\{\widetilde{P}_n\}$  can be obtained by the
Gram-Schmidt orthogonalization of the basis $\{\omega^n\}_{n=0}^\infty$ in $L^2({\Bbb R},
d\sigma)$, where
\begin{equation}\label{A.4}
 d\sigma(\omega):=\left\{
 \begin{array}{l}
  0\;\;\;\;\;\;\;\;\;\;\;\;\;\omega<a\\
  \varrho(\omega)\,d\omega\;\;\;\,a\leq\omega\leq b\\
  0\;\;\;\;\;\;\;\;\;\;\;\;\;\omega>b
 \end{array}\right.
\end{equation}
and $\varrho$ satisfies the Pearson equation with appropriately chosen polynomials
$(A(\omega),B(\omega))$. Namely if $deg\,B(\omega)=0\; (\textrm{i.e.}\;\;b_2=b_1=0)$ then we obtain
the Hermite polynomials; if $deg\,B(\omega)=1\; (\textrm{i.e.}\;\;b_2=0,\;b_1\neq 0)$, we obtain
the Laguerre polynomials and if $deg\,B(\omega)=2\; (\textrm{i.e.}\;\;b_2\neq 0)$, we obtain the
Jacobi polynomials. In the last case the boundary conditions (\ref{A.3}) hold if and only if $a$
and $b$ are roots of $B(\omega)$. For solution of the Pearson equation in these cases see table
\ref{tab:tableA.1.}. Additional conditions enforced on $A(\omega)$ by (\ref{A.3}) are presented in
this table too.

By straightforward calculation one can prove that the family of polynomials
$\left\{\frac{d^k}{d\omega^k}P_n(\omega)\right\}_{n=k}^\infty$, $k\in\Bbb N $, is orthogonal in the
space $L^2({\Bbb R},d\sigma^{(k)})$, where
\begin{equation}\label{A.5}
d\sigma^{(k)}(\omega):=B^k(\omega)\,d\sigma(\omega).
\end{equation}
The weight function $\varrho^{(k)}$ satisfies the Pearson equation on the interval $(a,b)$
associated with $(A^{(k)}(\omega),B(\omega))$, where
\begin{equation}\label{A.6}
A^{(k)}(\omega):=A(\omega)+k\frac{dB(\omega)}{d\omega}.
\end{equation}
\begin{proposition}

For a given Pearson data i.e. a pair $(A(\omega),B(\omega))$ on $(a,b)\subset\Bbb R$ the following
statements are equivalent:
\begin{enumerate}[A.]
 \item $\{P_n(\omega)\}_{n=0}^\infty$ form an orthonormal system
     in $L^2({\Bbb R}, d\sigma)$.
 \item  The polynomials are given by Rodrigues' formula
     \begin{equation}\label{A.7}
     P_n(\omega)=c_n\frac{1}{\varrho(\omega)}\cdot\frac{d^n}{d\omega^n}\left(\varrho(\omega)
     B^n(\omega)\right),
     \end{equation}
     where $c_n$ is the normalising constant. (see table \ref{tab:tableA.2.})
 \item  The polynomials $\{P_n(\omega)\}_{n=0}^\infty$ satisfy the
     differential equation
     \begin{equation}\label{A.8}
     \left(A(\omega)\frac{d}{d\omega}+B(\omega)\frac{d^2}{d\omega^2}\right)
     P_n(\omega)=\lambda_nP_n(\omega),
     \end{equation}
     where $\lambda_n=a_1n+b_2n(n-1)$.
 \item  The polynomials $\{P_n(\omega)\}_{n=0}^\infty$ are
     related by the tree-term recurrence formula (for  $h(n)$ and $b(n)$ see  table \ref{tab:tableA.3.})
     \begin{equation}\label{A.10}
     \omega P_n(\omega)=h(n)P_n(\omega)+b(n)P_{n-1}(\omega)+b(n+1)P_{n+1}(\omega)
     \end{equation}
with the initial condition
     \begin{equation}
     P_0(\omega)\equiv const= \displaystyle{\left[\int
d\sigma(\omega)\right]^ {-\frac{1}{2} }}.
     \end{equation}
\end{enumerate}
\end{proposition}

\begin{table*}
\caption{\label{tab:tableA.1.}}
\begin{ruledtabular}\renewcommand{\arraystretch}{2.2}
\begin{tabular}{llclc}
 \multicolumn{3}{c}{Pearson data}& \multicolumn{1}{c}{Weight function}& Additionally conitions\\\cline{1-3}
 $A(\omega)$&$B(\omega)$&$(a,b)$&\multicolumn{1}{c}{$\varrho(\omega)$}&\\ \hline
$A^{Her}(\omega)=a_1\omega+a_0$&$B^{Her}(\omega)=b_0$&$(-\infty,\infty)$&$\varrho^{Her}(\omega)=C
    e^{\frac{a_1}{2b_0}(\omega+\frac{a_0}{a_1})^2}$&$C>0,\;\;
   \frac{a_1}{b_0}<0$\\\hline
   $A^{Lag}(\omega)=a_1\omega+a_0$&$B^{Lag}(\omega)=b_1\omega+b_0$&$(-\frac{b_0}{b_1},\infty)
    $&$\varrho^{Lag}(\omega)=C(
    \omega+\frac{b_0}{b_1})^{\mu-1}    e^{\frac{a_1}{b_1}\omega}$&$C>0,\;\;\frac{a_1}{b_1}<0,
    $   \\&&&&$ \mu:=\frac{a_0b_1-b_0a_1}{b_1^2}>0  $\\\hline
  &&&&$C>0,\;a<b,\;\;b_2>0$,\\
 $ A^{Jac}(\omega)=a_1\omega+a_0$&$B^{Jac}(\omega)=b_2(\omega-a)(b-\omega)$&$(a,b)$
 &$\varrho^{Jac}(\omega)=C(\omega-a)^{\mu-1}(b-\omega)^{\nu-1}$&$\mu:=\frac{aa_1+a_0}{b_2(b-a)}>0$,\\
  &&&&$\;\nu:=
    \frac{ba_1+a_0}{b_2(a-b)}>0$\\
\end{tabular}
\end{ruledtabular}
\end{table*}

\begin{table}
\caption{\label{tab:tableA.2.}}
\begin{tabular}{l}\hline\hline\\
$c_n\textrm{--- in Rodrigues' formula} $\\\\\hline\\ $c_n^{Her}=\left(C
    n!\,(-a_1b_0)^n\sqrt{-\pi\frac{2b_0}{a_1}}\right)^{-\frac{1}{2}}$\\\\
    \hline\\
    $c_n^{Lag}=\left(Cn!\,(-a_1b_1)^n\left(-\frac{b_1}{a_1}\right)^{\mu+n}\Gamma(\mu+n)e
    ^{-\frac{a_1b_0} {b_1^2}}\right)^{-\frac{1}{2}}$\\\\\hline \\
$c_n^{Jac}=\left(Cn!\,b_2^{2n}(b-a)^{\mu+\nu+2n-1}\frac{\Gamma(\mu+n)\Gamma(\nu+n)}{(\mu+\nu+2n-1)\Gamma(\mu+\nu+n-1)}\right)
    ^{-\frac{1}{2}}$\\\\\hline\hline
\end{tabular}
\end{table}

\begin{table*}
\caption{\label{tab:tableA.3.}}
\renewcommand{\arraystretch}{2.2}
\begin{tabular}{ll}\hline\hline
  \multicolumn{1}{c}{$b(n)$}& \multicolumn{1}{c}{$h(n)$}\\\hline
 $b^{Her}(n)=\sqrt{-\frac{b_0}{a_1}n}$&
$h^{Her}(n)=-\frac{a_0}{a_1}$\\\hline
 $b^{Lag}(n)=-\frac{b_1}{a_1}\sqrt{n(n+\mu-1)}$&
$h^{Lag}(n)=-\frac{b_1}{a_1}\left(2n+\mu\right)-\frac{b_0}{b_1}$\\\hline
$b^{Jac}(n)=(b-a)\sqrt{\frac{n(\mu+n-1)(\nu+n-1)(\mu+\nu+n-2)}
  {(\mu+\nu+2n-3)(\mu+\nu+2n-2)^2(\mu+\nu+2n-1)}}$&
   $h^{Jac}(n)=\frac{2n(a+b)(\mu+\nu-1)+2n^2(a+b) -2b\mu-2a\nu+\mu
\nu (a+b)+b\mu^2+a\nu^2} {(\mu+\nu+2n-2)(\mu+\nu+2n)}$ \\\hline\hline
\end{tabular}
\renewcommand{\arraystretch}{0.1}
 {\scriptsize{ \bf {\underline{ Attention }}}} {\scriptsize{The
necessary condition $b(0)=0$ is automatically satisfied with the exception of the Jacobi case
for\newline $\mu=\nu=\frac{1}{2}$ and $\mu=\nu=\frac{3}{2}$ where we must put it additionally.}
\renewcommand{\arraystretch}{1}}
\end{table*}


\end{document}